\documentclass[amsmath,showpacs,nofootinbib,12pt]{revtex4-2}
\usepackage{graphicx}
\usepackage{dcolumn}
\usepackage{bm}
\usepackage{color} 
\usepackage{slashed}
\begin{document}
\newcommand{\hs}{\hspace*{0.5cm}}
\newcommand{\vs}{\vspace*{0.5cm}}
\newcommand{\be}{\begin{equation}}
\newcommand{\ee}{\end{equation}}
\newcommand{\bea}{\begin{eqnarray}}
\newcommand{\eea}{\end{eqnarray}}
\newcommand{\ben}{\begin{enumerate}}
\newcommand{\een}{\end{enumerate}}
\newcommand{\bde}{\begin{widetext}}
\newcommand{\ede}{\end{widetext}}
\newcommand{\nn}{\nonumber}
\newcommand{\crn}{\nonumber \\}
\newcommand{\Tr}{\mathrm{Tr}}
\newcommand{\non}{\nonumber}
\newcommand{\noi}{\noindent}
\newcommand{\al}{\alpha}
\newcommand{\la}{\lambda}
\newcommand{\bet}{\beta}
\newcommand{\ga}{\gamma}
\newcommand{\va}{\varphi}
\newcommand{\om}{\omega}
\newcommand{\pa}{\partial}
\newcommand{\+}{\dagger}
\newcommand{\fr}{\frac}
\newcommand{\bc}{\begin{center}}
\newcommand{\ec}{\end{center}}
\newcommand{\Ga}{\Gamma}
\newcommand{\de}{\delta}
\newcommand{\De}{\Delta}
\newcommand{\ep}{\epsilon}
\newcommand{\varep}{\varepsilon}
\newcommand{\ka}{\kappa}
\newcommand{\La}{\Lambda}
\newcommand{\si}{\sigma}
\newcommand{\Si}{\Sigma}
\newcommand{\ta}{\tau}
\newcommand{\up}{\upsilon}
\newcommand{\Up}{\Upsilon}
\newcommand{\ze}{\zeta}
\newcommand{\ps}{\psi}
\newcommand{\Ps}{\Psi}
\newcommand{\ph}{\phi}
\newcommand{\vph}{\varphi}
\newcommand{\Ph}{\Phi}
\newcommand{\Om}{\Omega}
\newcommand{\AdrHEPC}{Phenikaa Institute for Advanced Study and Faculty of Basic Science, Phenikaa University, Yen Nghia, Ha Dong, Hanoi 100000, Vietnam}

\title{Scotogenic model from an extended electroweak symmetry} 

\author{Phung Van Dong} 
\email{dong.phungvan@phenikaa-uni.edu.vn (corresponding author)}
\author{Duong Van Loi}
\email{loi.duongvan@phenikaa-uni.edu.vn}
\affiliation{\AdrHEPC} 

\date{\today}

\begin{abstract}

We argue that the higher weak isospin $SU(3)_L$ manifestly unifies dark matter and normal matter in its isomultiplets for which dark matter carries a conserved dark charge while normal matter does not. The resultant gauge symmetry is given by $SU(3)_C\otimes SU(3)_L \otimes U(1)_X\otimes U(1)_G$, where the first factor is the color group, while the rest defines a theory of scotoelectroweak in which $X$ and $G$ determine electric charge $Q=T_3-1/\sqrt{3}T_8+X$ and dark charge $D=-2/\sqrt{3}T_8+G$. This setup provides both appropriate scotogenic neutrino masses and dark matter stability as preserved by a residual dark parity $P_D=(-1)^D$. Interpretation of the dark charge is further discussed, given that $SU(3)_L$ is broken at very high energy scale.           

\end{abstract} 

\maketitle

\section{Introduction}

Neutrino mass \cite{Kajita:2016cak,McDonald:2016ixn} and dark matter \cite{Bertone:2004pz,Arcadi:2017kky} are the important questions in science which require the new physics beyond the standard model. Additionally, the standard model cannot address the quantization of electric charge and the existence of just three fermion families, as observed in the nature. 

Among attempts to solve these issues, the model based on $SU(3)_C\otimes SU(3)_L\otimes U(1)_X$ (called 3-3-1) gauge symmetry is well-motivated as it predicts the family number to be that of colors by anomaly cancellation \cite{331v1,331v2,331pp,331f,331flt}. Further, the charge quantization naturally arises in the 3-3-1 model for typical fermion contents \cite{ecq1,ecq2,ecq3,ecq4,ecq5}. The 3-3-1 model may supply small neutrino masses by implementing radiative and/or seesaw mechanisms  \cite{neu1,neu2,neu3,neu4,neu5,neu6,neu7,neu8,neu9,neu10,neu11,neu12,neu13} and dark matter stability by interpreting global/discrete symmetries \cite{d1,d2,d3,d4,d5,d6,d7,d8,d9,d10,d11,d12}. Recently, the 3-3-1 model may give a suitable solution to the $W$-mass anomaly \cite{wm331}.  

In the 3-3-1 model, the baryon minus lepton number $B-L$ generically neither commutes nor closes algebraically with $SU(3)_L$. This enlarges the 3-3-1 group to a complete gauge symmetry $SU(3)_C\otimes SU(3)_L\otimes U(1)_X\otimes U(1)_N$ (called 3-3-1-1) in which the last factor $N$ relates to $B-L$ via a $SU(3)_L$ charge and this setup reveals matter parity as a residual gauge symmetry \cite{3311v1,3311v2}. This matter parity stabilizes various dark matter candidates besides related phenomena as studied in \cite{3311dm1, 3311dm2, 3311dm3, 3311dm4}. The 3-3-1-1 model typically supplies neutrino masses via canonical seesaw, as suppressed by heavy right-handed neutrinos that exist due to anomaly cancellation and gain large Majorana masses from $N$-charge breaking. However, it may alternatively generate neutrino masses via scotogenic mechanism due to the existence of matter parity \cite{3311n1,3311n2,3311n3,3311n4,3311n5}. The cosmological inflation, asymmetric matter production, new abelian $N$-charge breaking, and effect of kinetic mixing between two $U(1)$ groups are extensively investigated in \cite{3311km1,3311km2,3311inf1,3311inf2,3311inf3,3311inf4} too.

The 3-3-1 symmetry has a property that unifies dark matter and normal matter in $SU(3)_L$ multiplets and normally couples dark matter in pairs in interactions \cite{3311v1}. Above, $B-L$ is realized in such a way that dark matter carries a wrong $B-L$ number opposite to that defined in the standard model for normal matter. Hence, dark matter is odd, governed by the matter parity. Since both dark matter and normal matter have $B-L$ charge, this setup implies a strict couple between the two kinds of matter through $B-L$ gauge portal. This work does not further examine such interacting effects of dark matter, especially under experimental detection \cite{3311dm1, 3311dm2, 3311dm3, 3311dm4}. Instead, we propose a dark charge for dark matter, while normal matter has no dark charge, which has a nature completely different from $B-L$ and relaxes such interaction. This interpretation of dark charge supplies naturally scotogenic neutrino mass and dark matter~\cite{sgsetup}, because the mentioned canonical seesaw including its right-handed neutrinos manifestly disappears.

A global version for dark charge under consideration was first discussed in \cite{d5} in attempt to find a mechanism for dark matter stability in 3-3-1 model and further promoted in~\cite{3311v1}. As electric charge $Q$ is unified with weak isospin $T_i$ $(i=1,2,3)$ in electroweak theory $SU(2)_L\otimes U(1)_Y$ for which $Q=T_3+Y$, the present proposal combines both electric charge $Q$ and dark charge $D$ in a higher weak isospin $T_n$ $(n=1,2,3,\cdots,8)$ yielding $SU(3)_L\otimes U(1)_X\otimes U(1)_G$ for which $Q=T_3+\beta T_8+X$ and $D=\beta' T_8+G$. Here the coefficients $\beta,\beta'$ determine the electric charge and dark charge of dark fields, respectively. This theory indeed unifies dark force and electroweak force in the same manner the electroweak theory does so for electromagnetic force and weak force, thus it is called scotoelectroweak, where ``scoto'' means darkness.  

The rest of this work is organized as follows. In Sec. \ref{pro} we propose the scotoelectroweak model. In Sec. \ref{sg} we examine scalar and gauge boson mass spectra. In Sec. \ref{neu} we obtain the scheme of neutrino mass generation. In Sec. \ref{dark} we investigate dark matter observables. In Sec. \ref{cst} we constrain the model and deliver a numerical investigation. In Sec. \ref{darkcharge} we give a realization of dark charge that the model refers to. Finally, we summarize our results and conclude this work in Sec. \ref{con}.     

\section{\label{pro} Scotoelectroweak setup}

In the standard model, the weak isospin $SU(2)_L$ arranges left-handed fermions in isodoublets $(\nu_{aL},e_{aL})\sim 2$ and $(u_{aL},d_{aL})\sim 2$, while putting relevant right-handed fermions in isosinglets $e_{aR}\sim 1$, $u_{aR}\sim 1$, and $d_{aR}\sim 1$, where $a=1,2,3$ is a family index. 

The standard model cannot explain nonzero neutrino masses and flavor mixing required by oscillation experiments. Additionally, it cannot explain the existence of dark matter which makes up most of the mass of galaxies and galaxy clusters. 

We argue that both the questions may be solved by existence of dark fields, a new kind of particles, which are  assumed, possessing a conserved dark charge $(D)$, normalized to unity for brevity, i.e. $D=\pm 1$. The content of dark fields and relevant dark symmetry are determined by enlarging the weak isospin $SU(2)_L$ to a higher symmetry, $SU(3)_L$. 

The fundamental representations of $SU(3)_L$ are decomposed as $3=2\oplus 1$ and $3^*=2^*\oplus 1$ under $SU(2)_L$. Hence, enlarging known fermion isodoublets ($2/2^*$) implies dark fermion isosinglets (1's) lying at the bottom of $3/3^*$, such as   
\bea \psi_{aL} = \begin{pmatrix}\nu_{aL}\\
e_{aL}\\
N_{aL}
\end{pmatrix}\sim 3,\hs
Q_{\al L} = \begin{pmatrix}d_{\al L}\\
-u_{\al L}\\
D_{\al L}
\end{pmatrix}\sim 3^*,\hs
Q_{3 L} = \begin{pmatrix} u_{3L}\\
d_{3L}\\
U_{3L}
\end{pmatrix}\sim 3,\eea where $\al=1,2$ is a family index as $a=1,2,3$ is. Furthermore, the relevant right-handed partners transform as $SU(3)_L$ singlets, 
\be e_{aR}\sim 1,\hs N_{aR}\sim 1,\hs u_{aR}\sim 1,\hs d_{aR}\sim 1,\hs D_{\al R}\sim 1,\hs U_{3R}\sim 1. \ee 

Above, the $[SU(3)_L]^3$ anomaly cancelation requires the third quark family (as well as those of leptons) transforming differently from the first two quark families \cite{ano1,ano2,ano3,ano4}. This condition demands that the number of fermion families matches that of color. As stated, $N_a$ and $U_3$ have a dark charge $D=1$, while $D_\al$ possesses a dark charge $D=-1$, as all collected in Tab.~\ref{tab1}. It is noted that all normal fields carry no dark charge, i.e. $D=0$.\footnote{As the standard model, the hypothetical right-handed neutrinos $\nu_{aR}$ are a gauge singlet having neither electric charge nor dark charge and are thus not imposed; whereas, the other right-handed fermions must be present, as already included.} We further assume $N_a$, $D_\al$, and $U_3$ possessing an electric charge $Q=0$, $-1/3$, and $2/3$ respectively like those of the 3-3-1 model with right-handed neutrinos.\footnote{Additionally, these dark leptons and quarks have the same $B,L$ numbers as usual leptons and quarks, hence $B$ and $L$ are global charges commuting with $SU(3)_L$ like those in the standard model, opposite to the original 3-3-1-1 model.} 

\begin{table}[h]
\bc
\begin{tabular}{lccccccccccccccccccccccc}
\\ \hline\hline 
Particle & $\nu_a$ & $e_a$ & $N_a$ & $u_a$ & $d_a$ & $D_\al$ & $U_3$ & $\eta_{1,2}$& $\rho_{1,2}$ & $\chi_3$ & $\eta_3$ & $\rho_3$ & $\chi_{1,2}$ & $\xi$ & $\phi$ & gluon & $\ga$ & $Z$ & $Z'$ & $Z''$ & $W$ & $X^0$ & $Y^-$ \\
\hline 
$D$ & 0 & 0 & 1 & 0 & 0 & $-1$ & $1$ & 0 & 0 & 0 & $1$& $1$ & $-1$ & 1 & $-2$ & 0 & 0 & 0 & 0 & 0 & 0 & $-1$ & $-1$\\
\hline
$P_D$ & $+$ & $+$ & $-$ & $+$ & $+$ & $-$ &$-$ & $+$ &$+$ &$+$ &$-$ &$-$ &$-$ &$-$ & $+$ &$+$ &$+$ &$+$ &$+$ &$+$ & $+$ & $-$ &$-$ \\
\hline\hline
\end{tabular}
\caption[]{\label{tab1} Dark charge ($D$) and dark parity ($P_D$) of the model particles.}
\ec
\end{table}

It is clear that $Q=\mathrm{diag}(0,-1,0)$ and $D=\mathrm{diag}(0,0,1)$ for lepton triplet $\psi_L$ which both neither commute nor close algebraically with $SU(3)_L$ charges. By symmetry principles, we obtain two new abelian charges $X$ and $G$ which complete the gauge symmetry,
\be SU(3)_C\otimes SU(3)_L \otimes U(1)_X\otimes U(1)_G, \ee
called 3-3-1-1, where $SU(3)_C$ is the color group, $SU(3)_L$ is previously given, while $X,G$ determine electric and dark charges, respectively, 
 \be Q=T_3-\fr{1}{\sqrt{3}}T_8+X,\hs D=-\fr{2}{\sqrt{3}}T_8+G,\label{qd}\ee               
where $T_{n}$ ($n=1,2,3,\cdots,8$) is $SU(3)_L$ charge. 

The fermion representation content under the 3-3-1-1 symmetry is given by 
\bea && \psi_{aL}\sim (1,3,-1/3,1/3),\hs Q_{\al L}\sim (3,3^*,0,-1/3), \hs Q_{3L}\sim (3,3,1/3,1/3),\\
&& e_{aR}\sim (1,1,-1,0),\hs N_{aR}\sim (1,1,0,1),\hs u_{aR}\sim (3,1,2/3,0),\\ 
&&d_{aR}\sim (3,1,-1/3,0),\hs D_{\al R}\sim (3,1,-1/3,-1),\hs U_{3R}\sim (3,1,2/3,1).\eea All the anomalies vanish. Indeed, since the 3-3-1 model is well established, it is sufficient to verify those associated with $U(1)_G$. 
\bea  [SU(3)_C]^2 U(1)_G  &\sim& \sum_{\mathrm{quarks}} (G_{q_L}-G_{q_R})\crn
&=& 2.3.(-1/3)+3.(1/3)-2.(-1)-1=0,
\eea
\bea
[SU(3)_L]^2 U(1)_G &\sim& \sum_{\mathrm{(anti)triplets}} G_{F_L}\crn
&=& 3.(1/3)+2.3.(-1/3)+3.(1/3)=0,
\eea
\bea [\mathrm{Gravity}]^2 U(1)_G &\sim& \sum_{\mathrm{fermions}} (G_{f_L}-G_{f_R})
 \crn
 &=& 3.3.(1/3)+2.3.3.(-1/3)+3.3.(1/3)\crn
 && -3.1-2.3.(-1)-3.1=0,
 \eea 
 \bea
 [U(1)_X]^2 U(1)_G &=& \sum_{\mathrm{fermions}}(X^2_{f_L}G_{f_L}-X^2_{f_R}G_{f_R})\crn
 &=& 3.3.(-1/3)^2.(1/3)+3.3.(1/3)^2(1/3)\crn
 &&-2.3.(-1/3)^2.(-1)-3.(2/3)^2.(1)=0,\eea
\bea
U(1)_X [U(1)_G]^2 &=& \sum_{\mathrm{fermions}}(X_{f_L}G^2_{f_L}-X_{f_R}G^2_{f_R})\crn
 &=& 3.3.(-1/3).(1/3)^2+3.3.(1/3)(1/3)^2\crn
 &&-2.3.(-1/3).(-1)^2-3.(2/3).(1)^2=0,\eea
 \bea [U(1)_G]^3&=& \sum_{\mathrm{fermions}}(G^3_{f_L}-G^3_{f_R})\crn
 &=&3.3.(1/3)^3+2.3.3.(-1/3)^3+3.3.(1/3)^3\crn
 &&-3.(1)^3-2.3.(-1)^3-3.(1)^3=0.\eea

The 3-3-1-1 symmetry breaking and mass generation are appropriately induced by
\bea \eta &=& \begin{pmatrix} \eta^0_1 \\
\eta^-_2 \\ \eta^0_3 \end{pmatrix} \sim (1,3,-1/3,1/3),\\ 
\rho &=& \begin{pmatrix} \rho^+_1 \\
\rho^0_2 \\ \rho^+_3 \end{pmatrix} \sim (1,3,2/3,1/3),\\ 
\chi &=& \begin{pmatrix} \chi^0_1 \\
\chi^-_2 \\ \chi^0_3 \end{pmatrix} \sim (1,3,-1/3,-2/3),\\
\phi &\sim& (1,1,0,-2),\hs \xi \sim (1,1,0,1).\eea Here $\phi$ couples to $N_R N_R$, breaks $U(1)_G$, and defines a dark parity. The fields $\eta$, $\rho$, and $\chi$ couple a fermion (anti)triplet to right-handed partners of the first, second, and third components respectively and break the 3-3-1 symmetry. The scalar $\xi$ analogous to a field in \cite{3311n4} couples to $\eta^\dagger \chi$ and $\phi$ inducing neutrino mass. Dark charge for scalars is included to Tab. \ref{tab1} too. Note that dark scalars include $\eta_3$, $\rho_3$, $\chi_{1,2}$, $\xi$, and $\phi$, which have $D\neq 0$, whereas the rest fields, $\eta_{1,2}$, $\rho_{1,2}$, and $\chi_3$, are normal scalars possessing $D=0$. 

Scalar fields develop vacuum expectation values (VEVs), such as 
\bea \langle \eta\rangle  &=& \begin{pmatrix} \fr{u}{\sqrt{2}} \\
0 \\ 0 \end{pmatrix},\hs  
\langle \rho\rangle  = \begin{pmatrix} 0 \\
\fr{v}{\sqrt{2}} \\ 0 \end{pmatrix},\hs 
\langle \chi\rangle  = \begin{pmatrix} 0 \\
0 \\ \fr{w}{\sqrt{2}} \end{pmatrix}, \hs
\langle \phi\rangle  = \fr{\La}{\sqrt{2}},\hs \langle \xi \rangle =0.\eea
The scheme of symmetry breaking is given by 
\bc
\begin{tabular}{c}
$SU(3)_C\otimes SU(3)_L \otimes U(1)_X\otimes U(1)_G$\\
$\downarrow \La, w $\\
$SU(3)_C\otimes SU(2)_L \otimes U(1)_Y \otimes P_D$\\
$\downarrow u,v$\\
$SU(3)_C\otimes U(1)_Q\otimes P_D$
\end{tabular}
\ec Here we assume $\La, w\gg u,v$ for consistency with the standard model. Besides the residual electric and color charges, the model conserves a residual dark parity, 
\be P_D=(-1)^{D}=(-1)^{-\fr{2}{\sqrt{3}}T_8+G}.\label{dp}\ee 

Indeed, a residual charge resulting from $SU(3)_L\otimes U(1)_X\otimes U(1)_G$ breaking must take the form $R=x_n T_n + y X + z G$. $R$ must annihilate the vacua $\langle \eta,\rho,\chi\rangle $, i.e. $R \langle \eta,\rho,\chi\rangle=0$, leading to $x_1=x_2=x_4=x_5=x_6=x_7=0$, $x_3=y$, and $x_8=-\fr{1}{\sqrt{3}}(y+2z)$. Substituting these $x$'s we get $R=yQ+zD$, where $Q,D$ are given as in (\ref{qd}). Obviously, $Q$ and $D$ commute, i.e. $[Q,D]=0$, implying that they are separated as two abelian subgroups. Additionally, $Q$ annihilates the vacuum $\langle \phi\rangle $, i.e. $Q\langle \phi\rangle =0$, implying that $Q$ is a final residual charge, conserved after breaking. For the remainder, $D$ is broken by $\langle \phi\rangle$, since $D\langle \phi\rangle=-2\La/\sqrt{2}\neq 0$. However, a residual symmetry of it, i.e. $P_D=e^{i\omega D}$, may be survived, i.e. $P_D \langle \phi\rangle = \langle \phi\rangle$, or $e^{i\omega (-2)}=1$, where $\omega$ is a transformation parameter. It leads to $\omega=k \pi$, for $k$ integer. Hence, $P_D=e^{i k \pi D}=(-1)^{kD}=\{1,(-1)^D\}\cong Z_2$ for which we redefine $P_D=(-1)^D$ to be dark parity as in (\ref{dp}). The dark parity (odd/even) of particles are collected in Tab. \ref{tab1} too. It is stressed that $\eta^0_3$, $\chi^0_1$, and $\xi$ do not have a nonzero VEV due to dark parity conservation.

We now write the total Lagrangian of the model, 
\be \mathcal{L}=\mathcal{L}_{\mathrm{kin}}+\mathcal{L}_{\mathrm{Yuk}}-V.\ee The kinetic part takes the form,   
\bea \mathcal{L}_{\mathrm{kin}}&=&\sum_F\bar{F}i\ga^\mu D_\mu F + \sum_S(D^\mu S)^\dagger (D_\mu S) -\fr{1}{4}\sum_A A_{\mu\nu}A^{\mu\nu},\eea where $F$, $S$, and $A$ denote fermion, scalar, and gauge-boson multiplets respectively, the covariant derivative $D_\mu$ and field strength tensors $A_{\mu\nu}$ are explicitly given by \bea D_\mu &=& \pa_\mu + ig_s t_n G_{n\mu} + ig T_n A_{n\mu} + ig_X X B_\mu + i g_G G C_\mu,\\ G_{n\mu\nu}&=&\pa_\mu G_{n\nu} -\pa_\nu G_{n\mu} -g_s f_{nmp}G_{m\mu}G_{p\nu},\\
A_{n\mu\nu}&=&\pa_\mu A_{n\nu} -\pa_\nu A_{n\mu} - g f_{nmp}A_{m\mu} A_{p\nu},\\
B_{\mu\nu}&=&\pa_\mu B_\nu -\pa_\nu B_\mu,\hs C_{\mu\nu}=\pa_\mu C_\nu -\pa_\nu C_\mu,\eea  where ($g_s,\ g,\ g_X,\ g_G$), ($G_{n\mu}, A_{n\mu}, B_\mu, C_\mu$), and ($t_n,\ T_n,\ X,\ G$) indicate coupling constants, gauge bosons, and charges according to 3-3-1-1 subgroups, respectively. Notice that all gauge bosons have $D=0$ behaving as normal fields, except for $X^0,Y^-$ coupled to $T_{4,5,6,7}$ having $D=-1$ and acting as dark vectors, which are all listed to Tab. \ref{tab1} too.   

The Yukawa Lagrangian is easily obtained,  
\bea \mathcal{L}_{\mathrm{Yuk}}&=&h^e_{ab}\bar{\psi}_{aL}\rho e_{bR} +h^N_{ab}\bar{\psi}_{aL}\chi N_{bR}+\fr 1 2 h'^N_{ab}\bar{N}^c_{aR} N_{bR}\phi \crn
&& + h^d_{\al a} \bar{Q}_{\al L}\eta^* d_{aR} +h^u_{\al a } \bar{Q}_{\al L}\rho^* u_{aR} + h^D_{\al \beta}\bar{Q}_{\al L} \chi^* D_{\beta R}\crn
&&+ h^u_{3a} \bar{Q}_{3L}\eta u_{aR}+h^d_{3a}\bar{Q}_{3L}\rho d_{aR}+ h^U_{33}\bar{Q}_{3L}\chi U_{3R} +H.c..\eea The scalar potential can be decomposed, 
\be V = V(\rho,\chi,\eta,\phi)+V(\xi), \ee where the first part relates to a potential that induces breaking,
\bea
V(\rho,\chi,\eta,\phi) &=& \mu^2_1\rho^\dagger \rho + \mu^2_2 \chi^\dagger \chi + \mu^2_3 \eta^\dagger \eta + \la_1 (\rho^\dagger \rho)^2 + \la_2 (\chi^\dagger \chi)^2 + \la_3 (\eta^\dagger \eta)^2\crn
&&+ \la_4 (\rho^\dagger \rho)(\chi^\dagger \chi) +\la_5 (\rho^\dagger \rho)(\eta^\dagger \eta)+\la_6 (\chi^\dagger \chi)(\eta^\dagger \eta)\crn
&& +\la_7 (\rho^\dagger \chi)(\chi^\dagger \rho) +\la_8 (\rho^\dagger \eta)(\eta^\dagger \rho)+\la_9 (\chi^\dagger \eta)(\eta^\dagger \chi) + (f\epsilon^{ijk}\eta_i\rho_j\chi_k+H.c.) \crn
&& + \mu^2 \phi^\dagger \phi + \la (\phi^\dagger \phi)^2 +\la_{10} (\phi^\dagger \phi)(\rho^\dagger\rho)+\la_{11}(\phi^\dagger \phi)(\chi^\dagger \chi)+\la_{12}(\phi^\dagger \phi)(\eta^\dagger \eta),\eea while the last part relates to a dark sector that induces neutrino mass,
\bea
V(\xi)&=&  \mu^2_\xi \xi^\dagger \xi + \la_\xi (\xi^\dagger \xi)^2  +\la_{13} (\xi^\dagger \xi)(\rho^\dagger\rho)+\la_{14}(\xi^\dagger \xi)(\chi^\dagger \chi) +\la_{15}(\xi^\dagger \xi)(\eta^\dagger \eta)\crn
&&+\la_{16}(\xi^\dagger \xi)(\phi^\dagger \phi)  + ( f_1 \phi \xi \xi +f_2 \xi \eta^\dagger \chi + \la_{17} \phi^*\xi^*\eta^\dagger \chi  +H.c.).
\eea Above, $h$'s and $\la$'s are dimensionless, while $\mu$'s and $f$'s have a mass dimension. We can consider the parameters $f$, $f_{1,2}$, and $\la_{17}$ to be real by absorbing their phases (if any) into appropriate scalar fields $\eta$, $\rho$, $\chi$, $\phi$, and $\xi$. That said, the potential conserves CP. We also suppose that CP is not broken by vacua, i.e. the VEVs $u$, $v$, $w$, and $\La$ are all real too. It is further noted that there are neither mixing between a scalar (CP-even) and a pseudo-scalar (CP-odd) due to CP conservation nor mixing between a $P_D$-even field and a $P_D$-odd field due to dark parity conservation.     

\section{\label{sg} Scalar and gauge boson masses}

\subsection{Scalar mass spectrum}

The potential $V(\rho,\chi,\eta,\phi)$ has been explicitly examined in \cite{3311dm1}. Let us summarize its result. First, expand the scalar fields around their VEVs,
\bea
\eta =
 \left(\begin{array}{c}
  \frac{u}{\sqrt{2}}\\
  0\\
  0
\end{array}\right)+ \left(\begin{array}{c}
   \frac{S_1+iA_1}{\sqrt{2}} \\
   \eta^-_2\\
   \frac{S^\prime_3+iA^\prime_3}{\sqrt{2}} \\
\end{array}\right), \hs 
\rho =
 \left(\begin{array}{c}
  0\\
  \frac{v}{\sqrt{2}}\\
  0
\end{array}\right)+ \left(\begin{array}{c}
 \rho^+_1 \\
  \frac{S_2+iA_2}{\sqrt{2}} \\
   \rho^+_3
\end{array}\right), \label{scl1}
\eea
\bea
\chi =
 \left(\begin{array}{c}
  0\\
  0\\
  \frac{w}{\sqrt{2}}\\
\end{array}\right)+ \left(\begin{array}{c}
   \frac{S^\prime_1+iA^\prime_1}{\sqrt{2}} \\
   \chi^-_2\\
   \frac{S_3+iA_3}{\sqrt{2}} 
\end{array}\right), \hs 
\phi = \frac{\La}{\sqrt{2}} + \frac{S_4+iA_4}{\sqrt{2}}, \label{scl2}
\eea and notice that the following approximations ``$\simeq$'' are given up to $(u,v)/(-f,w,\La)$ order. The usual Higgs field ($H$) and three new neutral scalars ($H_{1,2,3}$) are obtained by  
\bea && H \simeq \fr{ u S_1+v S_2}{\sqrt{u^2+v^2}},\hs H_1 \simeq \fr{-v S_1 + u S_2}{\sqrt{u^2+v^2}},\\
&& H_2 \simeq c_\varphi S_3-s_\varphi S_4,\hs H_3 \simeq s_\varphi S_3+c_\varphi S_4, \eea with mixing angle $t_{2\varphi}=\fr{\la_{11}w\La}{\la \La^2 -\la_2 w^2}$. The usual Higgs mass is appropriately achieved at the weak scale $m_H\sim (u,v)$, while the new scalar masses are
\bea && m^2_{H_1} \simeq -\fr{fw}{\sqrt{2}}\left(\fr{u}{v}+\fr{v}{u}\right),\\
&&m^2_{H_{2,3}} \simeq  \la_2 w^2 +\la \La^2 \mp \sqrt{(\la_2 w^2 -\la \La^2)^2 +\la^2_{11} w^2\La^2}.\eea 

A massive pseudo-scalar with corresponding mass is identified as \be \mathcal{A}=\fr{vw A_1+uw A_2+uv A_3}{\sqrt{u^2v^2+v^2w^2+u^2w^2}},\hs m^2_{\mathcal{A}}=-\fr{f}{\sqrt{2}}\left(\fr{vw}{u}+\fr{uw}{v}+\fr{uv}{w}\right).\ee Two charged scalars are given by 
\be H^\pm_4=\fr{v\chi^\pm_2+w \rho^\pm_3}{\sqrt{v^2+w^2}},\hs H^\pm_5=\fr{v\eta^\pm_2+u\rho^\pm_1}{\sqrt{u^2+v^2}},\ee with respective masses,
\be m^2_{H_4}=\left(\fr{\la_7}{2}-\fr{fu}{\sqrt{2}vw}\right)(v^2+w^2),\hs m^2_{H_5}=\left(\fr{\la_8}{2}-\fr{fw}{\sqrt{2}vu}\right)(v^2+u^2).\ee
A neutral complex scalar with corresponding mass is 
\be H'^0 \equiv \fr{S'+i A'}{\sqrt{2}}=\fr{u\chi^{0*}_1+w\eta^0_3}{\sqrt{u^2+w^2}},\hs m^2_{H'}=\left(\fr{\la_9}{2}-\fr{fv}{\sqrt{2}uw}\right)(u^2+w^2),\ee where the real $S'=(wS'_3+uS'_1)/\sqrt{u^2+w^2}$ and imaginary $A'=(wA'_3-uA'_1)/\sqrt{u^2+w^2}$ parts of $H'$ are degenerate with the same $H'$ mass. 

Except for the usual Higgs mass, all new scalar masses are given at $(w,\La,-f)$ scale. For the remaining fields, the massless Goldstone bosons of neutral gauge fields $Z$, $Z'$, and $Z''$ are identified as
\bea G_Z= \fr{u A_1- v A_2}{\sqrt{u^2+v^2}},\hs G_{Z'}=\fr{w(u^2+v^2)A_3-uv(vA_1+uA_2)}{\sqrt{(u^2+v^2)(u^2v^2+v^2w^2+u^2w^2)}},\hs G_{Z''}=A_4,\eea while those of charged/complex gauge fields $W^\pm$, $Y^\pm$, and $X^0$ take the form,
\be G^\pm_W=\fr{u\eta^\pm_2-v\rho^\pm_1}{\sqrt{u^2+v^2}},\hs G^\pm_Y=\fr{w\chi^\pm_2-v\rho^\pm_3}{\sqrt{v^2+w^2}},\hs G^0_X=\fr{w\chi^0_1-u\eta^{0*}_3}{\sqrt{u^2+w^2}}.\ee  

Because $\langle \xi\rangle =0$, the potential $V(\xi)$ does not affect the minimum conditions derived from $V(\rho,\chi,\eta,\phi)$ as in \cite{3311dm1}. In other words, $u,v,w,\La$ are uniquely given, assuming that $\mu^2<0$, $\mu^2_{1,2,3}<0$, $\la>0$, $\la_{1,2,3}>0$, and necessary conditions for $\la_{4,5,\cdots,12}$. Additionally, conservations of dark parity and electric charge imply that the presence of $\xi$, i.e. $V(\xi)$, modifies only the mass spectrum of $H'$ and $G_X$, or exactly $S'$ and $A'$, which includes
\bea V &\supset& \fr 1 2 \begin{pmatrix} S' & S'_5 \end{pmatrix} 
\begin{pmatrix} m^2_{H'} & \left(\fr{f_2}{\sqrt{2}}+\fr{\la_{17}\La}{2}\right)\sqrt{u^2+w^2} \\
\left(\fr{f_2}{\sqrt{2}}+\fr{\la_{17}\La}{2}\right)\sqrt{u^2+w^2} &  m^2_\xi+\sqrt{2}f_1\La\end{pmatrix} 
 \begin{pmatrix} S' \\
 S'_5\end{pmatrix}\crn
 &&+\fr 1 2\begin{pmatrix} A' & A'_5 \end{pmatrix}
 \begin{pmatrix} m^2_{H'} & \left(\fr{f_2}{\sqrt{2}}-\fr{\la_{17}\La}{2}\right)\sqrt{u^2+w^2} \\
 \left(\fr{f_2}{\sqrt{2}}-\fr{\la_{17}\La}{2}\right)\sqrt{u^2+w^2} & m^2_\xi-\sqrt{2}f_1\La \end{pmatrix} \begin{pmatrix} A' \\
 A'_5\end{pmatrix},\eea where $\xi\equiv (S'_5+iA'_5)/\sqrt{2}$ and $m^2_\xi\equiv \mu^2_\xi +\la_{13}v^2/2+\la_{14}w^2/2+\la_{15}u^2/2+\la_{16}\La^2/2$. Defining two mixing angles
\be t_{2\theta_R}=\fr{(\sqrt{2}f_2+\la_{17}\La)\sqrt{u^2+w^2}}{m^2_\xi+\sqrt{2}f_1\La-m^2_{H'}},\hs t_{2\theta_I}=\fr{(\sqrt{2}f_2-\la_{17}\La)\sqrt{u^2+w^2}}{m^2_\xi-\sqrt{2}f_1\La-m^2_{H'}},\ee
we obtain physical fields
\bea && R_1=c_{\theta_R} S'-s_{\theta_R} S'_5,\hs R_2=s_{\theta_R} S'+c_{\theta_R} S'_5,\\ 
&& I_1=c_{\theta_I} A'-s_{\theta_I} A'_5,\hs I_2=s_{\theta_I} A'+c_{\theta_I} A'_5,\eea 
with respective masses
\bea m^2_{R_{1,2}} &=&\fr 1 2 \left[ m^2_{H'}+m^2_\xi+\sqrt{2}f_1\La\right.\crn
&&\left. \mp \sqrt{(m^2_{H'}-m^2_\xi-\sqrt{2}f_1\La)^2+(\sqrt{2}f_2+\la_{17}\La)^2(u^2+w^2)}\right],\\
 m^2_{I_{1,2}} &=&\fr 1 2 \left[ m^2_{H'}+m^2_\xi-\sqrt{2}f_1\La \right.\crn
 &&\left. \mp\sqrt{(m^2_{H'}-m^2_\xi+\sqrt{2}f_1\La)^2+(\sqrt{2}f_2-\la_{17}\La)^2(u^2+w^2)}\right].\eea 

\subsection{Gauge boson mass spectrum}

The gauge bosons obtain mass from $\mathcal{L}\supset \sum_S(D^\mu \langle S\rangle)^\dagger (D_\mu \langle S\rangle)$. Substituting the VEVs, we get physical non-Hermitian gauge bosons
\be W^\pm_\mu = \fr{A_{1\mu}\mp i A_{2\mu}}{\sqrt{2}},\hs X^{0,0*}=\fr{A_{4\mu}\mp i A_{5\mu}}{\sqrt{2}},\hs Y^\mp = \fr{A_{6\mu}\mp i A_{7\mu}}{\sqrt{2}},\ee with respective masses,
\be m^2_W=\fr{g^2}{4}(u^2+v^2),\hs m^2_{X}=\fr{g^2}{4}(u^2+w^2),\hs m^2_Y=\fr{g^2}{4}(v^2+w^2).\ee $W$ is identical to that of the standard model and $u^2+v^2=(246\ \mathrm{GeV})^2$. 

Neutral gauge bosons are identified as 
\bea && A_\mu = s_W A_{3\mu}+c_W\left(-\fr{t_W}{\sqrt{3}}A_{8\mu}+\sqrt{1-\fr{t^2_W}{3}}B_\mu \right),\\
&&Z_\mu = c_W A_{3\mu}-s_W\left(-\fr{t_W}{\sqrt{3}}A_{8\mu}+\sqrt{1-\fr{t^2_W}{3}}B_\mu \right),\\
&&\mathcal{Z}'_\mu = \sqrt{1-\fr{t^2_W}{3}}A_{8\mu}+\fr{t_W}{\sqrt{3}}B_\mu,\eea where $s_W=e/g=\sqrt{3} t_X/\sqrt{3+4t^2_X}$, with $t_X=g_X/g$, is the sine of the Weinberg angle. The photon $A_\mu$ is massless and decoupled. The $Z$ boson that is identical to that of the standard model is radically lighter than the $\mathcal{Z}'$ boson of 3-3-1 model and the $C$ boson of $U(1)_G$. Although $Z$ mixes with $\mathcal{Z}'$ and $C$, at $(u,v)/(w,\La)$ order the field $Z$ is decoupled as a physical field possessing a mass, 
\be m^2_Z\simeq \fr{g^2}{4c^2_W}(u^2+v^2).\ee 

There remains a mixing between $\mathcal{Z}'$ and $C$, yielding physical fields by diagonalization,
\be Z'=c_\theta \mathcal{Z}'-s_\theta C,\hs Z''=s_\theta \mathcal{Z}'+c_\theta C,\ee with mixing angle and respective masses,
\bea t_{2\theta} &=& \fr{4\sqrt{3+t^2_X}t_Gw^2}{4t^2_G(w^2+9\La^2) - (3+t^2_X)w^2},\label{mixingtheta}\\ 
 m^2_{Z',Z''} &=& \fr{g^2}{18}\left\{4t^2_G(w^2+9\La^2)+(3+t^2_X)w^2\right.\crn
 &&\left.\mp\sqrt{[4t^2_G(w^2+9\La^2)-(3+t^2_X)w^2]^2+16(3+t^2_X)t^2_G w^4}\right\},\label{zpzppmass}\eea where $t_G=g_G/g$.
 
The above result is similar to that in \cite{3311dm1} since the scalar multiplets have a dark charge value equal to that for $B-L$. The difference would be explicitly in the couplings of $Z',Z''$ with matter fields because the normal fermions have $B-L$ while do not have dark charge. For comparison and further usage, we compute in Tab. \ref{tab2} the couplings of $Z'$ with fermions, while those for $Z''$ can be obtained from $Z'$ by replacing $c_\theta\rightarrow s_\theta$ and $s_
\theta\rightarrow -c_\theta$.

\begin{table}[h]
\bc
\begin{tabular}{lcc}
\hline\hline
$f$ & $g^{Z'}_V(f)$ & $g^{Z'}_A(f)$ \\
\hline  
$\nu_a $ & $\fr{c_\theta c_{2W}}{2\sqrt{3-4s^2_W}} -\fr 1 3 s_\theta c_W t_G$ & $\fr{c_\theta c_{2W}}{2\sqrt{3-4s^2_W}} -\fr 1 3 s_\theta c_W t_G$ \\
$e_a$ & $\fr{c_\theta (1-4 s^2_W) }{2\sqrt{3-4s^2_W}} -\fr 1 3 s_\theta c_W t_G$ & $\fr{c_\theta }{2\sqrt{3-4s^2_W}} -\fr 1 3 s_\theta c_W t_G$ \\ 
$N_a$ & $-\fr{c_\theta c^2_W}{\sqrt{3-4s^2_W}}-\fr 4 3 s_\theta c_W t_G$ & $-\fr{c_\theta c^2_W}{\sqrt{3-4s^2_W}}+\fr 2 3 s_\theta c_W t_G$ \\ 
$u_\al$ & $-\fr{c_\theta (3-8s^2_W)}{6\sqrt{3-4s^2_W}}+\fr 1 3 s_\theta c_W t_G$ & $- \fr{c_\theta }{2\sqrt{3-4s^2_W}} +\fr 1 3 s_\theta c_W t_G$ \\ 
$u_3$  & $\fr{c_\theta(3+2s^2_W)}{6\sqrt{3-4s^2_W}}-\fr 1 3 s_\theta c_W t_G$ & $\fr{c_\theta c_{2W}}{2\sqrt{3-4s^2_W}} -\fr 1 3 s_\theta c_W t_G$ \\ 
$d_\al$ & $-\fr{c_\theta (3-2s^2_W)}{6\sqrt{3-4s^2_W}}+\fr 1 3 s_\theta c_W t_G$ & $-\fr{c_\theta c_{2W}}{2\sqrt{3-4s^2_W}} + \fr 1 3 s_\theta c_W t_G$\\
$d_3$ & $ \fr{c_\theta \sqrt{3-4s^2_W}}{6}-\fr 1 3 s_\theta c_W t_G$ & $\fr{c_\theta }{2\sqrt{3-4s^2_W}} -\fr 1 3 s_\theta c_W t_G$ \\ 
$U$ & $-\fr{c_\theta (3-7s^2_W)}{3\sqrt{3-4s^2_W}}-\fr 4 3 s_\theta c_W t_G$ & $-\fr{c_\theta c^2_W}{\sqrt{3-4s^2_W}}+\fr 2 3 s_\theta c_W t_G$ \\
$D_\al$ & $\fr{c_\theta (3-5s^2_W)}{3\sqrt{3-4s^2_W}}+\fr 4 3 s_\theta c_W t_G$ & $\fr{c_\theta c^2_W}{\sqrt{3-4s^2_W}}-\fr 2 3 s_\theta c_W t_G$ \\
\hline\hline
\end{tabular}
\caption{\label{tab2} Couplings of $Z'$ with fermions; additionally, notice that $Z''$-fermion couplings derived from this table with replacement $c_\theta\rightarrow s_\theta$ and $s_\theta\rightarrow -c_\theta$. }
\ec
\end{table}

\section{\label{neu} Neutrino mass}
 
In the 3-3-1-1 model by gauging $B-L$, the right-handed neutrinos are required for anomaly cancellation. Consequently, neutrinos obtain a small mass via canonical seesaw mechanism, suppressed by large right-handed neutrino mass scales relating to $B-L$ breaking. In this kind of model, ordinary lepton doublets may couple to a scalar and fermions that both are odd under the matter parity, revealing an interesting possibility for scotogenic neutrino mass generation alternative to the above canonical seesaw \cite{3311n1,3311n2,3311n3,3311n4,3311n5}. The issue raised is how to suppress this canonical seesaw since the $B-L$ breaking scale is not necessarily large for the latter. The most studies have chosen $B-L$ charges for right-handed neutrinos to be $-4,-4,+5$ which avoid their coupling to usual leptons and Higgs boson. But one must introduce two scalar singlets coupled to these right-handed neutrinos in order to make them appropriately heavy, hence expressing a complicated $U(1)_N$ Higgs sector with two unreasonable pseudo Nambu-Goldstone bosons. Additionally, the fermions that are odd under the matter parity responsible for the mentioned scotogenic setup are not necessarily present under the theoretical ground, unlike the unwanted $\nu_{aR}$. The present 3-3-1-1 model by gauging dark charge properly overcomes such issues. Indeed, $\nu_{aR}$ are not required by dark charge anomaly cancellation, thus the canonical seesaw disappears. Additionally, $N_{aR}$ must be present for dark charge anomaly cancellation, which are odd under dark parity and coupled to usual leptons via a scalar triplet. We introduce only an extra scalar singlet $\xi$ that necessarily separates the relevant $H'$ (i.e. $S',A'$) mass, yielding a neutrino mass generation scheme to be more economical than the previous studies. 

First note that charged leptons and every (usual and exotic) quarks gain appropriate masses from the Yukawa Lagrangian, as usual/similar to the 3-3-1 model. Neutral fermions obtain a mass matrix of form, 
\be \mathcal{L}_{\mathrm{Yuk}} \supset -\fr 1 2  \begin{pmatrix} \bar{N}_{aL} & \bar{N}^c_{aR} \end{pmatrix} \begin{pmatrix} 0 & m^D_{ab} \\ 
m^D_{ba} & m^R_{ab}\end{pmatrix}\begin{pmatrix} N^c_{bL} \\ 
N_{bR}\end{pmatrix}+H.c.,
\ee where $m^D=-h^Nw/\sqrt{2}$ and $m^R=- h'^N\La/\sqrt{2}$ are Dirac and (right-handed) Majorana masses for $N$, respectively. We can diagonalize the generic mass matrix, yielding \be \mathcal{L}_{\mathrm{Yuk}}\supset -\fr 1 2 \bar{N}^c_k M_k N_k, \ee for $k=1,2,\cdots,6$, where $(N^c_{aL},N_{aR})=(U_{ak},V_{ak})N_k$ relates the gauge states to mass eigenstates $N_k$ with mass eigenvalues $M_k$.    

\begin{figure}[h]
\bc
\includegraphics[scale=0.75]{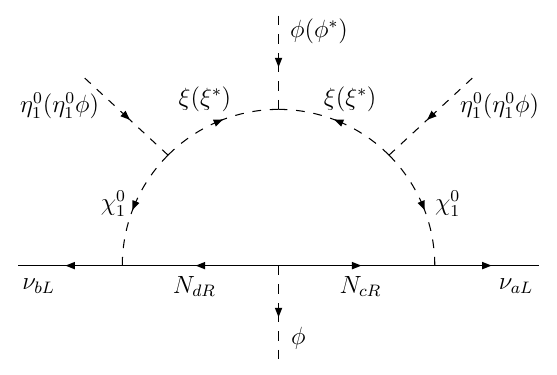}
\includegraphics[scale=0.75]{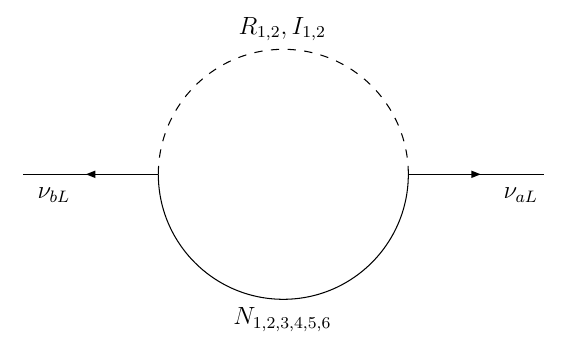}
\caption[]{\label{fig1} Neutrino mass generation in the scotoelectroweak theory, where left and right diagrams are given in flavor and mass eigenbases, respectively.}
\ec
\end{figure}

What concerns is neutrino mass generation Lagrangian which is collected from those in Yukawa interactions and scalar potential, such as
\bea \mathcal{L} &\supset& \fr{u h^N_{ab}V_{bk}}{\sqrt{2}\sqrt{u^2+w^2}}\bar{\nu}_{aL} (c_{\theta_R}R_1+s_{\theta_R} R_2-i c_{\theta_I} I_1-i s_{\theta_I}I_2) N_{k} \crn
&& +\fr{w h^N_{ab}V_{bk}}{\sqrt{u^2+w^2}}\bar{\nu}_{aL}G^0_X N_{k}  - \fr 1 2 M_k N^2_k+H.c.\crn
&&-\fr 1 2 m^2_{R_1} R^2_1-\fr 1 2 m^2_{R_2} R^2_2-\fr 1 2 m^2_{I_1} I^2_1-\fr 1 2 m^2_{I_2}I^2_2,\eea where we have used $\chi^0_1=(uH'^{0*}+wG^0_X)/\sqrt{u^2+w^2}=[u(c_{\theta_R}R_1+s_{\theta_R} R_2-i c_{\theta_I} I_1-i s_{\theta_I}I_2)/\sqrt{2}+wG^0_X]/\sqrt{u^2+w^2}$ and $N_{bR}=V_{bk} N_k$. Neutrino mass generation Feynman diagram is depicted in Fig. \ref{fig1} in both flavor basis (left panel) and mass eigenbasis (right panel).  
 Neutrino mass is induced in form of $\mathcal{L}\supset -\fr 1 2 \bar{\nu}_{aL} (m_\nu)_{ab} \nu^c_{bL}+H.c.$ in which 
\bea (m_\nu)_{ab} &=& \fr{u^2}{u^2+w^2}\fr{(h^N V)_{ak} (h^N V)_{bk} M_k}{32\pi^2}\crn
&& \times \left(\fr{c^2_{\theta_R}m^2_{R_1}\ln\fr{M^2_k}{m^2_{R_1}}}{M^2_k-m^2_{R_1}}-\fr{c^2_{\theta_I}m^2_{I_1}\ln\fr{M^2_k}{m^2_{I_1}}}{M^2_k-m^2_{I_1}}+\fr{s^2_{\theta_R}m^2_{R_2}\ln\fr{M^2_k}{m^2_{R_2}}}{M^2_k-m^2_{R_2}}-\fr{s^2_{\theta_I}m^2_{I_2}\ln\fr{M^2_k}{m^2_{I_2}}}{M^2_k-m^2_{I_2}}\right).\label{numdd} \eea

Remarks are in order
\ben 
\item The divergent one-loop contributions corresponding to $R_{1,2}$ and $I_{1,2}$ are cancelled out due to $c^2_{\theta_R}-c^2_{\theta_I}+s^2_{\theta_R}-s^2_{\theta_I}=0$.
\item For gauge realization of the dark parity (even the matter parity instead), the relevant inert scalar doublet $(\chi_1,\chi_2)$ may approximate as Goldstone mode of a gauge vector doublet $(X,Y)$, i.e. $(\chi_1,\chi_2)\sim (G_X,G_Y)$. Both $G_X$ and $X$ do not contribute to neutrino mass since they possess a degenerate mass between particle and antiparticle, opposite to its global versions \cite{sgsetup,vdoublet}. 
\item Contributing to neutrino mass is a scalar singlet $\eta_3$ that mixes with $\chi_1$, thus suppressed by $(u/w)^2\sim 10^{-3}$ besides the usual loop factor $(1/32\pi^2)\sim 10^{-3}$, another intermediate scalar singlet $\xi$ that connects to $\eta_3$, and the singlet mass splittings $\Delta m^2/m^2\sim f_1/\La\sim f_2\la_{17}/\La$ as well as Majorana masses $M_k \sim \La$ for $N_k$, all governed by dark charge breaking field $\langle \phi\rangle \sim \La$. It translates to \be m_\nu\sim \left(\fr{h^N}{10^{-2}}\right)^2\times \left(\fr{f_1,f_2\la_{17}}{\mathrm{GeV}}\right)\times 0.1\ \mathrm{eV},\ee appropriate to experiment, given that $h^N\sim 10^{-2}$, and the soft coupling $f_{1,2}\sim 1$ GeV is not necessarily small, in contrast to \cite{3311n4}. This is due to a double suppression between the weak and new physics scales, $(u/w)^2$.  
\een
      
\section{\label{dark} Dark matter}

Contributing to the scotogenic neutrino masses is two kinds of dark field, the dark scalars $R_{1,2},I_{1,2}$ and the dark fermions $N_{1,2,\cdots,6}$. In contrast to the 3-3-1-1 model by gauging $B-L$, the dark scalars in the present model are now separated in mass $m_{R_1}\neq m_{I_1}$ and $m_{R_2}\neq m_{I_2}$. This presents interesting coannihilation phenomena between $R_1$ and $I_1$ as well as $R_2$ and $I_2$ that set the relic density, if each of them is interpreted to be dark matter. Additionally, the dark scalar mass splitting would avoid dangerous scattering processes of $R_1/I_1$ or $R_2/I_2$ with nuclei in direct detection experiment due to mediators of $Z,Z',Z''$. The phenomenology of dark scalar candidates is quite analogous to those studied in the 3-3-1 model with inert multiplets \cite{d7,d10,d11}, which will be skipped. In what follows we assume the dark fermions containing dark matter, namely the dark matter candidate is assigned as $N_1$ which has a mass smaller than other $N$'s, dark scalars, and dark vectors. Therefore, this $N_1$ is absolutely stabilized by dark parity conservation. 

\begin{figure}[h]
\bc
\includegraphics[scale=1]{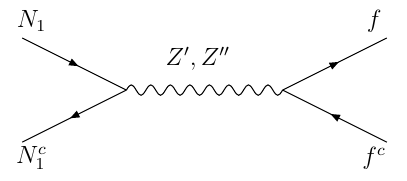}
\caption[]{\label{fig2} Fermion dark matter annihilation to normal matter.}
\ec
\end{figure}

A distinct feature between the 3-3-1-1 model by gauging $B-L$ and the 3-3-1-1 model by gauging dark charge is that $N_1$ in the former has $B-L=0$, while $N_1$ in the latter has $D=1\neq 0$. Therefore, in the present model $N_1=U^*_{a1}N^c_{aL}+V^*_{a1}N_{aR}$ has both (left and right) chiral couplings to $Z',Z''$, such as
\bea \mathcal{L} &\supset& -\left[\left(\fr{g c_W c_\theta}{\sqrt{3-4s^2_W}}+\fr{g_G s_\theta }{3}\right)U^*_{a1}U_{a1} -g_G s_\theta  V^*_{a1}V_{a1} \right] \bar{N}_{1} \ga^\mu N_{1} Z'_\mu\crn
&&-\left[\left(\fr{g c_W s_\theta}{\sqrt{3-4s^2_W}}-\fr{g_G c_\theta }{3}\right)U^*_{a1}U_{a1}+g_G c_\theta  V^*_{a1}V_{a1}  \right] \bar{N}_{1} \ga^\mu N_{1} Z''_\mu,\label{n1coupling} \eea where the terms $V_{a1}$ (exactly of $N_{aR}$) exist only in the present model, which sets the neutrino mass above. Specially, we will examine the effect of $N_{aR}$ by assuming $||V_{a1}||\gg ||U_{a1}||$, i.e. the dark matter $N_1\simeq V^*_{a1} N_{aR}$, up to the small term $U^*_{a1}N^c_{aL}$, to be most right-handed. Combined with unitarity condition, we have $V^*_{a1}V_{a1}=1-U^*_{a1}U_{a1}\simeq 1$ while $U^*_{a1}U_{a1}\simeq 0$, given at the leading order $||U_{a1}||$. Eq. (\ref{n1coupling}) becomes \be \mathcal{L} \supset g_G s_\theta   \bar{N}_{1} \ga^\mu N_{1} Z'_\mu-g_G c_\theta \bar{N}_{1} \ga^\mu N_{1} Z''_\mu. \ee In the early universe, $N_1$ annihilates to usual fields via $Z',Z''$ portals as in Fig.~\ref{fig2} which set the relic density. Here the $Z',Z''$ couplings with usual fermions $(f=\nu,e,u,d)$ can be found in Tab. \ref{tab2}. It is stressed that there are no $t$-channel annihilations exchanged by $X,Y$ dark vectors, in contrast to \cite{3311v1}. Additionally, the Higgs portal interactions of $N_1$ with normal matter are small and suppressed.       

The dark matter annihilation cross-section is computed as  
\bea \langle \sigma v\rangle_{N_1} = \fr{\langle v^2\rangle g^4 m^2_{N_1}}{12\pi c^4_W}\sum_{f,x,y}\fr{g^x_A(N_1)g^y_A(N_1)N_C(f) [g^x_V(f) g^y_V(f) + g^x_A(f) g^y_A(f)]}{(4m^2_{N_1}-m^2_x)(4m^2_{N_1}-m^2_y)},\eea where $x,y=Z',Z''$, $N_C(f)$ refers to the color number of $f$, and $g^{Z'}_A(N_1)=s_\theta c_W t_G$ and $g^{Z''}_A(N_1)=-c_\theta c_W t_G$ are given in the mass basis of $N$, as mentioned. The thermal average over dark matter relative velocity obeys $\langle v^2\rangle =6/x_F$ for $x_F=m_{N_1}/T_F \simeq 20$ at freeze-out temperature. Further, the dark matter relic density can be approximated as $\Om_{N_1} h^2 \simeq 0.1\ \mathrm{pb}/\langle \sigma v \rangle_{N_1} \simeq 0.12$, where the last value is given by experiment \cite{pdg}.

Because $N_1$ is a Majorana particle, it scatters with quarks in direct detection experiment only through spin-dependent (SD) effective interaction exchanged by $Z',Z''$ analogous to the diagram in Fig. \ref{fig2} for $f=q$, namely
\be \mathcal{L}_{\mathrm{eff}} \supset \fr{g^2}{4c^2_W}\sum_{q,x}\fr{g^x_A(N_1) g^x_A(q)}{m^2_x} (\bar{N}_1 \ga^\mu \ga_5 N_1)(\bar{q}\ga_\mu \ga_5q),\label{monojetdm}\ee where $g^x_A(N_1)$ and $g^x_A(q)$ for $x=Z',Z''$ have been given. The SD cross-section determining scattering of $N_1$ with a target neutron ($n$) is given by 
\bea \sigma^{\mathrm{SD}}_{N_1} = \fr{3g^4 m^2_n}{4\pi c^4_W} \sum_{x,y}\fr{g^x_A(N_1) g^y_A(N_1)[g^x_A(u)\la_u^n+g_A^x(d)(\la^n_d+\la^n_s)][g^y_A(u)\la_u^n+g_A^y(d)(\la^n_d+\la^n_s)]}{m^2_x m^2_y},\eea where $x,y=Z',Z''$, and the fractional quark-spin coefficients are $\la^n_u=-0.42$, $\la^n_d=0.85$, and $\la^n_s=-0.88$ for neutron \cite{dmsd}. Notice that dark matter scattering with proton leads to a similar bound, which is not of interest.   

\section{\label{cst} Constraining}

Because the neutrino masses are governed by $h^N$ and $f_{1,2},\la_{17}$ all independent of the gauge portal, the dark matter observables can appropriately be constrained to be independent with those for the neutrino.\footnote{Note that $N_1$ mass that enters dark matter observables can be induced by a $h'^N$ coupling. The other $h'^N$ and $h^N$ couplings are sufficient to recover neutrino data.} Only the supplemental conditions that are relevant to dark matter are the mass regime for WIMP stability, the collider limit for $Z',Z''$ and $X,Y$ masses, and FCNCs, which will be studied in order. 
 
\subsection{WIMP stability} 

It is easy to adjust relevant Yukawa couplings and scalar potential parameters so that $N_1$ is lighter than other dark fermions and dark scalars. But for dark vectors, we must impose\be m_{N_1}<m_{X,Y}\simeq \fr{g}{2}w,\ee where $m_{N_1} = M_1$ is the mass of $N_1$ as mentioned and the last approximation is given at the leading order $u,v\ll w$. 

\subsection{Collider bound} 

In our model, $Z'$ and $Z''$ couple to lepton and quark quite equally (cf. Tab. \ref{tab2}). Hence, the LEPII and LHC experiments would make similar bounds on these new gauge bosons, analogous to a sequential $Z'$ boson that has the same couplings as the standard model $Z$ boson~(see, e.g., \cite{lepii,lhc}). Besides $Z',Z''$, the 3-3-1-1 symmetry contains two new non-Hermitian gauge bosons $X,Y$. In contrast to a sequential $W'$ boson that possesses the same couplings as the usual $W$ boson, the gauge fields $X,Y$ are odd under dark parity and couple only to a dark fermion and a normal fermion (similarly for scalars and gauge bosons by themselves). Because of dark parity conservation, the dark fields like $X,Y$ must be produced in pairs in particle colliders, in contrast to $Z',Z''$ that may be singly created. It is necessary to consider the LEPII bound for dilepton signal and then investigate dark matter, dilepton, and dijet signals at the LHC.

\subsubsection{LEPII}

The LEPII experiment \cite{lepii} studied possesses $e^+e^-\to f\bar{f}$ for $f=\mu,\tau$, exchanged by new neutral gauge bosons as $Z',Z''$. Since the LEPII collision energy $\sqrt{s}=209$ GeV is much smaller than $Z',Z''$ masses, such processes can be best described by effective interactions, obtained by integrating $Z',Z''$ out, to be  
\be \mathcal{L}_{\mathrm{eff}}\supset \sum_{x}\fr{g^2}{c^2_W m^2_x }[\bar{e}\ga^\mu (a^x_L (e)P_L+a^x_R(e)P_R)e][\bar{f}\ga_\mu (a^x_L(f)P_L +a^x_R(f)P_R)f],\ee where we label $x=Z',Z''$. The chiral couplings defined by $a^x_{L,R}(f)=\fr 1 2 [g_V^x (f)\pm g_A^x (f)]$ can directly be extracted from Tab. \ref{tab2}. 

Since the charged leptons possess universal gauge couplings, we further write 
\be \mathcal{L}_{\mathrm{eff}}\supset \sum_{x}\fr{g^2 [a^x_L(e)]^2}{c^2_W m^2_x }(\bar{e}\ga^\mu P_L e)(\bar{f}\ga_\mu P_L f) + (LR)+(RL)+(RR),\label{chiint}\ee where the last three terms $(\cdots)$ differ from the first term only in chiral structures, where the concerning couplings are explicitly supplied by \be a^{Z'}_L(e)=\fr{c_\theta c_{2W}}{2\sqrt{3-4s^2_W}}-\fr 1 3 s_\theta c_W t_G,\hs a^{Z''}_L(e)=a^{Z'}_L(e)|_{c_\theta\to s_\theta,s_\theta\to-c_\theta}.\ee  

The LEPII experiment investigated the chiral interaction types in (\ref{chiint}), making several constraints on the effective couplings. They typically indicate to \cite{lepii1}
\be \sum_{x}\fr{g^2 [a^x_L(e)]^2}{c^2_W m^2_x } = \fr{g^2}{c^2_W}\left\{\fr{[a^{Z'}_L(e)]^2}{m^2_{Z'}}+\fr{[a^{Z''}_L(e)]^2}{m^2_{Z''}}\right\}<\fr{1}{(6\ \mathrm{TeV})^2}.\label{lepc}\ee

By the way, let us remind the reader that since the dark matter mass in our model is beyond the weak scale, the dark matter cannot be produced (on-shell) by heavy mediators $Z'/Z''$ or $X/Y$ at the LEPII, as kinematically forbidden.  

\subsubsection{LHC}

In contrast to $Z',Z''$ that can significantly decay to normal fields (as well as possible dark fields), the dark gauge bosons $X,Y$ only decay to a lighter dark field, such as a dark fermion $N,U,D$ or a dark scalar $H_4,H',R_{1,2},I_{1,2}$, due to dark parity conservation. Since $N_1$ dark matter mass is limited below the mass of the lightest (labelled $V$) of $X,Y$, we assume $V$ lighter than the remaining dark fermions and the dark scalars; hence, $V$ decays only to the dark matter. Since the LHC is indeed energetic, a pair of dark vectors may be produced as $pp\to VV^*$, followed by $V,V^*$ decays to $N_1$ dark matter, such as $V\to l N_1$ and $V^*\to l^c N_1$, where $l$ defines one of usual leptons $(\nu,e)$ that couples $V$ to $N_1$, $\mathcal{L}\supset -\fr{g}{\sqrt{2}}U^*_{l1}\bar{l} V\!\!\!\!/ P_L N^c_1+H.c.$. The LHC searches for dilepton signals $ll^c$ recoiled against large missing transverse energy ($E\!\!\!\!/_T$) carried by a pair of dark matter $N_1 N_1$. The dilepton cross section is 
\bea \sigma(pp \to ll^c+E\!\!\!\!/_T)&=&\sigma(pp\to VV^*\to ll^c N_1 N_1)\crn
&=&\sigma(pp \to VV^*)\times \mathrm{Br}(V\to l N_1)\times \mathrm{Br}(V^*\to l^c N_1), \eea with the help of narrow width approximation, where $\mathrm{Br}(V\to l N_1)=\mathrm{Br}(V^*\to l^c N_1)=1$, as given. The process $pp \to VV^*$ proceeds through $s$-channel contributions by $\ga,Z,Z',Z''$ and $t$-channel contributions by $U,D$, which conserves unitarity. However, the cross section $\sigma(pp\to VV^*)$ is dominantly governed by $\ga,Z$, because $V=(X^0,Y^-)$ transforms nontrivially under the electroweak symmetry as $(2,-1/2)$, whereas the new mediators $(Z',Z'')$ and $(U,D)$ only remove unphysical contributions coming from bad behavior of $V$ at high energy and are subdominant, given that all $Z',Z'',U,D$ are above 1 TeV (cf. \cite{ppvv}). Hence, the cross section is given at quark level as 
\be \sigma(qq^c\to VV^*)\simeq \fr{\pi \al^2}{36E^2}\left(1-\fr{m^2_V}{E^2}\right)^{3/2} \left[Q^2_q Q^2_V +\fr{Q_q Q_V v_q v_V}{s^2_W c^2_W}+\fr{(v^2_q+a^2_q) v^2_V}{s^4_W c^4_W} \right],\ee where the energy of incident quark is $E=\fr 1 2 \sqrt{s}>m_V\gg m_Z$. The $Z$-quark couplings are $v_q=T_{3q}-2s^2_W Q_q$ and $a_q=T_{3q}$, while the $Z$-$V$ coupling is $v_V=T_{3V}-s^2_W Q_V$. We denote $Q_{q,V}$ and $T_{3q,V}$ as electric charge and weak isospin of $q,V$, respectively. This cross section obeys the equivalence theorem, $\sigma(qq^c\to VV^*)\simeq \sigma(qq^c\to G_V G^*_V)$, where $G_V$ is the Goldstone boson associated with $V$; or in other words, $V$ is identical to $G_V$ at high energy. This longitudinal mode $G_V$ has the same statistic and gauge quantum numbers with a hypothetical left-handed slepton ($\tilde{l}$) in SUSY, i.e. $\sigma(qq^c\to VV^*)\simeq \sigma(qq^c\to \tilde{l}\tilde{l}^*)$. The LHC \cite{ppvv1} have studied slepton-pair production, then decaying to dilepton plus missing energy, i.e. $pp \to \tilde{l}\tilde{l}^*\to ll^c \tilde{\chi}^0_1\tilde{\chi}^0_1$, assuming $\mathrm{Br}(\tilde{l}\to l \tilde{\chi}^0_1)=1$, making a bound for charged slepton mass $m_{\tilde{l}}>700$ GeV. The SUSY result applies to our case without change, i.e. \be m_V>700\ \mathrm{GeV},\ \mathrm{or}\ w=\fr{2}{g}m_V >2.15\ \mathrm{TeV},\ee for $g=0.652$. That said, the equivalence theorem justifies high energy behavior of $V$ as a well-studied slepton, predicting its mass bound, as given.                       

The LHC searches for jet signals recoiled against large missing energy ($E\!\!\!\!/_T$) carried by a pair of dark matter, putting strong constraints on interactions between quarks and dark matter mediated by a new neutral gauge boson. In this model, both $Z',Z''$ contribute to the process, where notice that $m_{Z'}<m_{Z''}$. As will be seen, the $N_1$ dark matter observables are strictly set by one of $Z',Z''$ mass resonances, either $m_{N_1}=\fr 1 2 m_{Z'}$ or $m_{N_1}=\fr 1 2 m_{Z''}$. For the latter with $Z''$ resonance, $Z''$ decay to dark matter is strongly suppressed by a phase space factor $(1-4m^2_{N_1}/m^2_{Z''})^{3/2}\sim 10^{-3}$ since $Z''$ has purely axial-vector coupling to $N_1$, i.e. $g^{Z''}_V(N_1)=0$. Additionally, $Z'$ decay to dark matter is kinematically forbidden, because of $m_{Z'}<2m_{N_1}$. Since $g^{Z''}_A(N_1)=-c_\theta c_W t_G$ is similar in size to usual fermion couplings in Tab.~\ref{tab2}, the mono-jet cross section is proportional to \be \sigma(pp \to j+E\!\!\!\!/_T)\sim [(g^{Z''}_V(q))^2+(g^{Z''}_A(q))^2] (1-4m^2_{N_1}/m^2_{Z''})^{3/2}, \ee suppressed by $10^{-3}$, presenting a negligible signal strength (cf. \cite{monojetdms}). For the former with $Z'$ resonance, $Z'$ negligibly contributes to the mono-jet cross section, analogous to $Z''$ in the latter case. However, $Z''$ now decays to a pair of dark matter, because of $m_{Z''}>2m_{N_1}$. In this case, the mono-jet cross section is proportional to \be \sigma(pp \to E\!\!\!\!/_T+j)\sim (g^{Z''}_V(q))^2+(g^{Z''}_A(q))^2.\ee  Ref. \cite{monojetdms} used a simplified dark matter model, in which an axial vector mediator $Z_A$ couples to a Dirac dark matter $\chi$ by $g_{\chi}=1$ and universally to quarks by $g_q=1/4$, making a bound $m_{Z_A}>2$ ($1.5$) TeV for $m_\chi$ just above the weak scale (600 GeV) and relaxing for $m_\chi>600$~GeV. Assuming $t_G\sim 1\sim t_\theta$, this result is possibly applied to the present model without change, since the $Z''$ couplings to quarks and dark matter possess quite the same sizes as the simplified dark matter model. That said, the mono-jet search bounds $m_{Z''}>1.5$--2 TeV for $N_1$ mass beyond the weak scale but below 600 GeV, while it relaxes for $m_{N_1}>600$~GeV. Since the dark matter observables are necessarily governed by $Z'$ resonance demanding $m_{Z''}> 2m_{N_1}\simeq m_{Z'}$ beyond few TeV, the bound corresponding to the low dark matter mass regime $m_{N_1}<600$ GeV  does not apply. Thus, this kind of bound is automatically satisfied by dark matter physics, which need not be further imposed.

Alternative to the invisible decays to dark matter, $Z',Z''$ can effectively decay to standard model particles, giving rise to promising signals at the LHC, such as dilepton and dijet, examined in order. Since $Z'$ and $Z''$ interact with usual fermions similarly in strength, a search designed at the LHC that bounds $Z'$ does so for $Z''$, because of $m_{Z'}<m_{Z''}$. Notice that the LHC searches only for a single new neutral gauge boson. Hence, it is sufficient to study the LHC bound for $Z'$, while the $Z''$ mass is possibly separated from that of $Z'$. There are two alternative cases that make $Z''$ decoupled, either i) $\La\gg w$ that reduces the 3-3-1-1 model to the relevant 3-3-1 model whose $Z'$ bound is well-studied, or ii) $w\gg \La$ that reduces the 3-3-1-1 model to the standard model plus the $D$ dark charge whose interpretation will be further investigated in Sec. \ref{darkcharge}. There remains a generic case according to $w\sim \La$ for which the $Z'$-$Z''$ mixing is finite and dependent on $(w,\La)$. In this case, the $Z'$ bound must depend on this mixing, i.e. $(w,\La)$, but $Z''$ always obeys such bound, since $m_{Z'}<m_{Z''}$. 

The cross section that produces a final state of dilepton ($l\bar{l}$) or dijet ($u\bar{u}$, $d\bar{d}$) at the LHC via $Z'$ exchange can be evaluated by narrow width approximation, 
\be \sigma(pp \to Z'\to f\bar{f})=\fr 1 3 \sum _q \fr{d L_{q\bar{q}}}{dm^2_{Z'}}\hat{\sigma}(q\bar{q}\to Z')\mathrm{Br}(Z'\to f\bar{f}),\label{lhcff}\ee where we define $f=(l,u,d)$, and the luminosity $dL_{q\bar{q}}/dm^2_{Z'}$ can be obtained from \cite{luminosity} for the LHC with $\sqrt{s}=13$ TeV or higher energy if relevant. The partonic peak cross-section $\hat{\sigma}(q\bar{q}\to Z')$ and the branching decay ratio $\mathrm{Br}(Z'\to f\bar{f})=\Ga(Z'\to f\bar{f})/\sum_{f'} \Ga(Z'\to f'\bar{f}')$ are given, respectively, by 
\bea && \hat{\sigma}(q\bar{q}\to Z')=\fr{\pi g^2}{12 c^2_W} \left[(g^{Z'}_V(q))^2+(g^{Z'}_A(q))^2\right],\label{lhczp}\\
&& \Ga(Z'\to f'\bar{f'}) =\fr{g^2 m_{Z'}}{48\pi c^2_W} N_C(f')\left[(g^{Z'}_V(f'))^2+(g^{Z'}_A(f'))^2\right],\label{lhcdc}
\eea where we denote $f'$ to be all standard model fermions $(\nu,e,u,d)$, which contain the product $f$ and  neutrinos $\nu$. In the total width, we exclude decays $Z'\to N_1 N_1$ and other new particles, which mostly include dark fields heavier than $N_1$, which either do not significantly modify the signal strength or are kinematically suppressed. For each value of $\La$ as in Tab. \ref{tab3}, $w$ is extracted as a function of $m_{Z'}$ from (\ref{zpzppmass}). Substituting this $w$ to (\ref{mixingtheta}), the mixing angle $\theta$ is given as a function of $m_{Z'}$. Hence, demanding the dilepton cross section $\sigma(pp\to Z'\to l\bar{l})$ satisfies both the latest ATLAS \cite{atlascms1} and CMS \cite{atlascms2} constraints taking width per resonance mass to be 3\% and 0.6\%, respectively. We obtain a $Z'$ bound according to each $\La$, as collected in Tab. \ref{tab3}. This $Z'$ bound gives a corresponding $w$ value, as listed in Tab. \ref{tab3} too. We have used $s^2_W=0.231$, $\al=1/128$, and $t_G=g_G/g=1$. It is clear that when $\La$ is as large as 50 TeV, $m_{Z'}$ approaches a bound $4.133$ TeV close to that of the 3-3-1 model \cite{futurelhc}. Vice versa, when $w$ is as large as 50 TeV, $m_{Z'}$ tends to a bound $3.39$ TeV as the dark gauge boson, detailed below. Alternatively, demanding the dijet cross section $\sigma(pp \to Z' \to q \bar{q})$ obeys the latest ATLAS bound for $\sigma\times A\times \mathrm{Br}$ taking kinematic acceptance $A\simeq 0.4$ \cite{atlasdijetn}. Further, we need only compare the largest dijet cross section with experiment, which comes from the decay mode with $q=b$, i.e. $Z'\to b\bar{b}$. Hence, we find a $Z'$ mass limit corresponding to each value of $\La$, which subsequently translates to a relevant $w$ value, as all collected in Tab. \ref{tab4}. It is stressed that when $\La$ as large as 50 TeV, the $Z'$ mass approaches that limit of the 3-3-1 model, $m_{Z'}\simeq 1.3201$ TeV. Vice versa, when $\La$ is as small as 3.89 TeV, which is similar to $w$ size, it slightly modifies this 3-3-1 bound down to 1.2938 TeV, since the quark couplings to $Z'$ are not much sensitive to the $Z'$-$Z''$ mixing. Below $\La=3.89$ TeV, there is neither available data nor any bound for $Z'$ because the predicted dijet cross-section is negligible. Last, but not least, since the $Z'$-quark and $Z'$-lepton couplings have quite the same magnitude, as well as the current bound on dijet signals is less sensitive than that of dilepton signals, the lower bound for $Z'$ mass implied by the dijet search is quite smaller than that arising from the dilepton search, as given.         

\begin{table}[h]
\begin{tabular}{lcccccccccc}
\hline\hline
$\Lambda$ & {3.89}& {3.9} & {4} & {4.3} & {4.7} & {5} & {5.4} & {6} & {7} & {9} \\
\hline 
$m_{Z'}$ & {3.392} & {3.397} & {3.390} & {3.415} & {3.547} & {3.659} & {3.734} & {3.803} & {3.872} & {3.997}\\
$w$ & {50.098} & {47.163} & {24.442} & {15.92} & {14.214} & {13.774} & {12.921} & {12.088} & {11.37} & {10.974}\\
\hline\hline
$\Lambda$ & {11} & {13} & {15} & {17} & {19} & {23} & {27} & {31} & {35} & {50}\\
\hline
$m_{Z'}$ & {4.047} & {4.072} & {4.091} & {4.104} & {4.110} & {4.116} & {4.122} & {4.124} & {4.126} & {4.133} \\
$w$ & {10.774} & {10.662} & {10.605} & {10.571} & {10.539} & {10.495} & {10.476} & {10.46} & {10.45} & {10.441}\\
\hline\hline
\end{tabular}
\caption[]{\label{tab3} LHC dilepton bound for $Z'$ gauge boson mass according to each value of $\La$, where the relevant $w$ value is supplied with respective to the $Z'$ mass limit, where all values are given in TeV.}
\end{table}

\begin{table}[h]
\begin{tabular}{lccccccc}
\hline\hline
$\Lambda$ & 3.89  & 5 & 10 & 20 & 30 & 40 & 50\\
\hline
$m_{Z'}$ & 1.2938 & 1.2996 & 1.3158 &  1.3196 &  1.3200 & 1.3200 & 1.3201\\
$w$ &  3.406 & 3.360 & 3.337 & 3.331 & 3.329 & 3.328 & 3.328\\
\hline\hline
\end{tabular}
\caption[]{\label{tab4} LHC dijet bound for $Z'$ gauge boson mass corresponding to each value of $\La$, which yields a relevant value for $w$ too, where all values are defined in TeV.}
\end{table}              

The projected high-luminosity and high-energy LHC as well as the Future Circular Collider will make a stronger bound for $Z',Z''$ masses, if no positive signal for $Z',Z''$ is found. Since such future colliders supply the strongest limit among the others for $Z',Z''$, the dark matter physics governed by $Z',Z''$ interpreted below may be changed. However, this assumption (for negative $Z',Z''$ search and its implication) is indeed out of the scope of this work, a task to be published elsewhere. Here, let us attract the reader's attention to a detailed study on this matter in the relevant 3-3-1 model \cite{futurelhc}.              

\subsection{FCNC} 

\subsubsection{FCNC coupled to new neutral gauge bosons}

Since quark families transform differently under the gauge symmetry, there must be FCNCs coupled to $Z',Z''$. They arise from the gauge interaction, 
\be \mathcal{L}\supset -g \bar{F}\ga^\mu[T_3 A_{3\mu} + T_8 A_{8\mu}+ t_X (Q-T_3+T_8/\sqrt{3}) B_\mu + t_G (D+2 T_8/\sqrt{3}) C_\mu] F,\ee where we have substituted $X,G$ from (\ref{qd}). It is noted that all leptons and exotic quarks do not flavor change, while the couplings of $Q$, $T_3$, and $D$ always conserve flavors, due to dark parity conservation. What remains is only usual quarks coupled to $T_8$, 
\bea  \mathcal{L} &\supset& -g \bar{q}_L\ga^\mu T_{q8} q_L(A_{8\mu}+ t_X/\sqrt{3} B_\mu + 2t_G/\sqrt{3} C_\mu)\crn
&\supset& \bar{q}'_{iL}\ga^\mu q'_{jL}(V^*_{qL})_{3i} (V_{qL})_{3j} (g'Z'_\mu+g''Z''_\mu),\label{adttn}\eea which flavor changes for $i\neq j$ ($i,j=1,2,3$). Above, $q$ denotes either $u=(u_1,u_2,u_3)$ or $d=(d_1,d_2,d_3)$ whose $T_8$ value is $T_{q8}=\fr{1}{2\sqrt{3}}\mathrm{diag}(-1,-1,1)$. Additionally, $q'$ defines mass eigenstates, either $u'=(u,c,t)$ or $d'=(d,s,b)$, related to gauge states by $q_{L,R}=V_{qL,R}q'_{L,R}$ which diagonalizes relevant quark mass matrices. The $g',g''$ couplings are
\be g'= \fr 2 3 g_G s_\theta-\fr{gc_\theta  c_W}{\sqrt{3-4s^2_W}},\hs g''=g'(c_\theta\to s_\theta,s_\theta\to -c_\theta). \ee 

For convenience, we rewrite the couplings in (\ref{adttn}) as \be \mathcal{L}\supset \Theta^{Z'}_{ij} \bar{q}'_{iL}\ga^\mu q'_{jL}Z'_\mu +\Theta^{Z''}_{ij} \bar{q}'_{iL}\ga^\mu q'_{jL}Z''_\mu, \ee where $\Theta^{Z'}_{ij}=g'(V^*_{qL})_{3i} (V_{qL})_{3j}$ and $\Theta^{Z''}_{ij}=g''(V^*_{qL})_{3i} (V_{qL})_{3j}$. Integrating $Z',Z''$ out, we obtain an effective Hamiltonian contributing to the relevant meson mixing,
\be \mathcal{H}^G_{\mathrm{eff}} = (\bar{q}'_{iL}\ga^\mu q'_{jL})^2 \left[\fr{(\Theta^{Z'}_{ij})^2}{m^2_{Z'}}+\fr{(\Theta^{Z''}_{ij})^2}{m^2_{Z''}}\right]\sim \fr{1}{m^2_{Z',Z''}}.\label{Gcon} \ee Aligning the quark mixing to down quark sector, i.e. $V_{uL}=1$, it implies $V_{dL}=V_{\mathrm{CKM}}$. Given that the new physics effect dominantly arises from the above effective interaction, the existing data on neutral meson mixings $K^0$-$\bar{K}^0$ and $B^0_{d,s}$-$\bar{B}^0_{d,s}$ give quite the same bounds on the new physics. Indeed, the mixing systems $K^0$-$\bar{K}^0$, $B^0_d$-$\bar{B}^0_d$, and $B^0_s$-$\bar{B}^0_s$ constrain  
\bea && \fr{(\Theta^{Z'}_{12})^2}{m^2_{Z'}}+\fr{(\Theta^{Z''}_{12})^2}{m^2_{Z''}}<\fr{1}{(10^4\ \mathrm{TeV})^2},\\
&&\fr{(\Theta^{Z'}_{13})^2}{m^2_{Z'}}+\fr{(\Theta^{Z''}_{13})^2}{m^2_{Z''}}<\fr{1}{(500\ \mathrm{TeV})^2},\\
&&\fr{(\Theta^{Z'}_{23})^2}{m^2_{Z'}}+\fr{(\Theta^{Z''}_{23})^2}{m^2_{Z''}}<\fr{1}{(100\ \mathrm{TeV})^2}, \eea respectively \cite{nmmb}. The CKM elements are given by $(V_{dL})_{31}=0.00857$, $(V_{dL})_{32}=0.04110$, and $(V_{dL})_{33}=0.999118$ \cite{pdg}. This leads to \be \fr{g'^2}{m^2_{Z'}}+\fr{g''^2}{m^2_{Z''}} <\fr{1}{(3.52\ \mathrm{TeV})^2},\ \fr{1}{(4.28\ \mathrm{TeV})^2},\ \mathrm{and}\ \fr{1}{(4.11\ \mathrm{TeV})^2},\label{fcncc} \ee according to the above meson mixings, respectively. In what follows, a bound $4$ TeV is applied, i.e. $(g'/m_{Z'})^2+(g''/m_{Z''})^2<(1/4\ \mathrm{TeV})^2$, without loss of generality. 

Remarks are in order. i) If $\La\gg w$, $Z''$ is superheavy with mass $m_{Z''}\simeq 2g_G\La\simeq -3g''\La$, while $Z'$ obtains a mass $m_{Z'}\simeq gc_W w/\sqrt{3-4s^2_W}\simeq -g'w$ at $w$ scale, where note that the mixing angle $\theta\simeq 0$. The FCNC bound is translated to $w>4$ TeV, realizing a 3-3-1 symmetry at this energy, as usual (where $U(1)_G$ is decoupled). ii) If $w\gg \La$, $Z''$ is superheavy with mass at $w$ scale, while $Z'$ gets a mass at $\La$ scale. In this case, the mixing angle approaches $t_\theta\simeq \sqrt{3+t^2_X}/(2t_G)=3c_W/(2t_G \sqrt{3-4s^2_W})$, where $s_W=\sqrt{3}t_X/\sqrt{3+4t^2_X}$ is previously given, such that $g'\simeq 0$, while $g''\simeq -2g_G/(3c_\theta)\neq 0$. That said, $(g'/m_{Z'})^2+(g''/m_{Z''})^2\to 0$, implying that there are neither FCNC at this limit $w\gg \La$ nor a bound on $\La$, realizing a dark symmetry $U(1)_D$ with a potential light dark gauge boson (where the 3-3-1-1 symmetry is decoupled, broken down to the standard model and the dark charge).      

The FCNC may arise from interactions of fermions with scalars, potentially modifying the above result. In what follows, the contribution of scalars to FCNC is evaluated.  

\subsubsection{FCNC coupled to neutral scalars and pseudoscalars}

According to Tab. \ref{tab1}, the normal scalars which are $P_D$-even potentially couple to FCNC including two doublets $(\eta_1,\eta_2)$ and $(\rho_1,\rho_2)$ as well as two singlets $\chi_3$ and $\phi$. Notice that the dark scalars $\eta_3$, $\rho_3$, $\chi_{1,2}$ and $\xi$ are $P_D$-odd, not coupled to FCNC. Further, the contributions of $\chi_3$ and $\phi$ to FCNC are suppressed by $(u,v)/(w,\La)$ as compared to those by $\eta_{1,2}$ and $\rho_{1,2}$ and are thus negligible. On the other hand, the interactions of usual leptons with neutral scalars do not flavor change. Hence, the FCNC significantly comes from the couplings of usual quarks with the two scalar doublets, such as 
\bea \mathcal{L}_{\mathrm{Yuk}} &\supset& h^d_{\al a} \bar{Q}_{\al L} \eta^* d_{aR} + h^d_{3a} \bar{Q}_{3L} \rho d_{aR} + h^u_{\al a} \bar{Q}_{\al L} \rho^* u_{aR} + h^u_{3a} \bar{Q}_{3L}\eta u_{aR}+H.c.\crn
&\supset& h^d_{\al a} \bar{d}_{\al L} \fr{u+S_1-i A_1}{\sqrt{2}}d_{aR}+h^d_{3a}\bar{d}_{3L}\fr{v+S_2+iA_2}{\sqrt{2}}d_{aR}\crn
&&-h^u_{\al a} \bar{u}_{\al L}\fr{v+S_2-i A_2}{\sqrt{2}}u_{aR}+h^u_{3a}\bar{u}_{3L}\fr{u+S_1+iA_1}{\sqrt{2}}u_{aR}+H.c.\crn
&\supset &-\bar{q}_L m_q q_R + \bar{q}_L \Ga^H_q q_R H +\bar{q}_L \Ga^{H_1}_q q_R H_1+\bar{q}_L i\Ga^{\mathcal{A}}_q q_R\mathcal{A} + H.c.,\eea where $q$ is either $u=(u_1,u_2,u_3)$ or $d=(d_1,d_2,d_3)$, while the physical scalar fields $H$, $H_1$, and $\mathcal{A}$ are related to $S_{1,2}$ and $A_{1,2}$, such as 
\be \begin{pmatrix} H\\
H_1\end{pmatrix}\simeq \begin{pmatrix} c_\beta & s_\beta \\
-s_\beta & c_\beta \end{pmatrix}\begin{pmatrix} S_1\\
S_2\end{pmatrix},\hs \begin{pmatrix} \mathcal{A}\\
G_Z\end{pmatrix}\simeq \begin{pmatrix} c_\beta & s_\beta \\
-s_\beta & c_\beta \end{pmatrix}\begin{pmatrix} A_2\\
A_1\end{pmatrix},\ee where $t_\beta\equiv v/u$ [notice that the approximations are governed by $-f\sim (w,\La)\gg (u,v)$, as all supplied in the scalar sector above.] 

The mass matrices of down-type and up-type quarks are given by \bea &&(m_d)_{\al a}=-h^d_{\al a} u/\sqrt{2},\hs (m_d)_{3a}=-h^d_{3a} v/\sqrt{2},\\ 
&&(m_u)_{\al a}= h^u_{\al a} v/\sqrt{2},\hs (m_u)_{3a}=-h^u_{3a}u/\sqrt{2}.\eea Additionally, the couplings $\Ga$'s take the form,
\bea (\Ga^H_d)_{\al a} = h^d_{\al a} c_\beta/\sqrt{2},\hs (\Ga^H_d)_{3 a}=h^d_{3a} s_\beta/\sqrt{2},\\ 
(\Ga^H_u)_{\al a} = -h^u_{\al a} s_\beta/\sqrt{2},\hs (\Ga^H_u)_{3 a}=h^u_{3a} c_\beta/\sqrt{2},\eea from which the remaining couplings are followed, $\Ga^{H_1}_q=\Ga^H_q (c_\beta\to -s_\beta,s_\beta\to c_\beta)$, $\Ga^{\mathcal{A}}_d=\Ga^H_d (c_\beta\to -s_\beta,s_\beta\to c_\beta)$, and $\Ga^{\mathcal{A}}_u=\Ga^H_u (c_\beta\to s_\beta,s_\beta\to -c_\beta)$. It is clear that $\Ga^H_q=-m_q/v_{\mathrm{w}}$ for $q$ to be either up-type or down-type quarks, where $v_\mathrm{w}=\sqrt{u^2+v^2}=246$ GeV is the weak scale. Hence, there is no FCNC associated with the standard model Higgs field $H$, in contradiction to \cite{fcnchiggs}. Note that $m_q$ is diagonalized by $V^\dagger_{qL} m_q V_{qR}=m_{q'}$ to be a diagonal matrix of either up-type or down-type quark masses, where $V_{qL,R}$ and $q'$ were previously defined. It is straightforwardly to derive $m_q=V_{qL} m_{q'} V^\dagger_{qR}$, thus 
\bea && h^d_{\al a}=-\fr{\sqrt{2}}{u}(V_{dL} m_{d'} V^\dagger_{dR})_{\al a},\hs h^d_{3 a}=-\fr{\sqrt{2}}{v}(V_{dL} m_{d'} V^\dagger_{dR})_{3 a},\label{yuk1c}\\
&& h^u_{\al a}=\fr{\sqrt{2}}{v}(V_{uL} m_{u'} V^\dagger_{uR})_{\al a},\hs h^u_{3 a}=-\fr{\sqrt{2}}{u}(V_{uL} m_{u'} V^\dagger_{uR})_{3 a},\label{yuk2c} \eea used for determining $\Ga^{H_1}_q$ and $\Ga^{\mathcal{A}}_q$ as a function of ($V_{qR}$,$t_\beta$) since $V_{qL}$ is related to the CKM matrix as previously supposed.

The FCNC is coupled/governed only by $H_1,\mathcal{A}$, such as
\be \mathcal{L}_{\mathrm{Yuk}}\supset \Theta^{H_1}_{i j} \bar{q}'_{iL} q'_{jR}H_1+ i \Theta^{\mathcal{A}}_{ij}\bar{q}'_{iL} q'_{jR} \mathcal{A}+H.c.,\ee where $\Theta^{S}_{ij}=(V^\dagger_{qL} \Ga^{S}_q V_{qR})_{ij}$ for $S=H_1,\mathcal{A}$ (notice $i\neq j$). With the aid of unitarity conditions for $V_{qL}$ and $V_{qR}$ as well as relations (\ref{yuk1c}) and (\ref{yuk2c}) for Yukawa couplings, we derive 
\be \Theta^{H_1}_{ij}=\Theta^{\mathcal{A}}_{ij}=-\fr{v_{\mathrm{w}}m_{d'_j}}{uv}(V^*_{dL})_{3i}(V_{dL})_{3j},
 \ee for down quarks, while 
 \be -\Theta^{H_1}_{ij}=\Theta^{\mathcal{A}}_{ij}=-\fr{v_{\mathrm{w}}m_{u'_j}}{uv}(V^*_{uL})_{3i}(V_{uL})_{3j}, \ee for up quarks, which all are independent of $V_{qR}$, as expected. Integrating the heavy fields $H_1$ and $\mathcal{A}$ out we obtain an effective Hamiltonian,
\bea \mathcal{H}^S_{\mathrm{eff}} &=& -(\bar{q}'_{iL} q'_{jR})^2\left[\fr{(\Theta^{H_1}_{ij})^2}{m^2_{H_1}}-\fr{(\Theta^{\mathcal{A}}_{ij})^2}{m^2_{\mathcal{A}}}\right] - (\bar{q}'_{iR} q'_{jL})^2\left[\fr{(\Theta^{H_1 *}_{ji})^2}{m^2_{H_1}}-\fr{(\Theta^{\mathcal{A}*}_{ji})^2}{m^2_{\mathcal{A}}}\right]\crn
&&- 2(\bar{q}'_{iL} q'_{jR})(\bar{q}'_{iR} q'_{jL}) \left[\fr{\Theta^{H_1}_{ij}\Theta^{H_1 *}_{ji}}{m^2_{H_1}}-\fr{\Theta^{\mathcal{A}}_{ij}\Theta^{\mathcal{A}*}_{ji}}{m^2_{\mathcal{A}}}\right]\sim \fr{1}{m^2_{H_1}}-\fr{1}{m^2_{\mathcal{A}}},\label{Scon}\eea where the coefficient ``2'' arises from two equal contributions, $(LR)(RL)$ and $(RL)(LR)$. Because of $(u,v)\ll (-f,w,\La)$, the $H_1$ and $\mathcal{A}$ mass splitting is small, given at weak scale (cf. the scalar section above). That said, the scalar contribution $1/m^2_{H_1}-1/m^2_{\mathcal{A}}\sim (u,v)^2/f^2w^2$ is at order $(u,v)^2/w^2\sim 10^{-2}$--$10^{-3}$ small compared to that of the gauge contribution (\ref{Gcon}).

To see explicitly the strong suppression of scalar contribution, we consider the new physics contributions to neutral meson mixings $K^0$-$\bar{K}^0$ and $B^0_{d,s}$-$\bar{B}^0_{d,s}$ generically coming from the new neutral gauge (\ref{Gcon}) and the new neutral scalar (\ref{Scon}), such as
\bea \mathcal{H}_{\mathrm{eff}} &=& \mathcal{H}_{\mathrm{eff}}^G+\mathcal{H}^S_{\mathrm{eff}}\crn
&=& [(V_{qL})_{3i}(V_{qL})_{3j}]^2\left\{\left(\fr{g'^2}{m^2_{Z'}}+\fr{g''^2}{m^2_{Z''}}\right)(\bar{q}'_{iL}\ga^\mu q'_{jL})^2-\fr{v^2_{\mathrm{w}}}{u^2v^2}\left(\fr{1}{m^2_{H_1}}-\fr{1}{m^2_{\mathcal{A}}}\right)
\right.\crn
&&\left.\times \left[m^2_{q'_j}(\bar{q}'_{iL} q'_{jR})^2+m^2_{q'_i} (\bar{q}'_{iR} q'_{jL})^2+2 m_{q'_i} m_{q'_j}(\bar{q}'_{iL} q'_{jR})(\bar{q}'_{iR} q'_{jL})\right]\right\},\eea assuming the effective couplings to be real, without loss of generality. This yields the mass difference for $K^0$-$\bar{K}^0$ mixing system as
\be \Delta m_K = 2 \Re \langle K^0 | \mathcal{H}_{\mathrm{eff}}| \bar{K}^0\rangle,\ee where $(q'_i,q'_j)=(d,s)$. With the aid of the hadronic matrix elements \cite{nmmele}, i.e. 
\bea && \langle K^0| (\bar{d}_L \ga^\mu s_L)^2|\bar{K}^0\rangle = \fr 1 3 m_K f^2_K,\\
&& \langle K^0| (\bar{d}_L s_R)^2|\bar{K}^0\rangle = \langle K^0| (\bar{d}_R s_L)^2|\bar{K}^0\rangle = -\fr{5}{24}\left(\fr{m_K}{m_s+m_d}\right)^2m_K f^2_K,\\
&&\langle K^0| (\bar{d}_L s_R)(\bar{d}_R s_L)|\bar{K}^0\rangle = \left[\fr{1}{24}+\fr 1 4 \left(\fr{m_K}{m_s+m_d}\right)^2\right]m_K f^2_K,\eea we obtain
\bea \Delta m_K &\simeq & m_K f^2_K [(V_{dL})_{31}(V_{dL})_{32}]^2\crn
&&\times \left[\fr 2 3 \left(\fr{g'^2}{m^2_{Z'}}+\fr{g''^2}{m^2_{Z''}}\right)+\fr{v^2_{\mathrm{w}}m^2_K}{12u^2v^2} \left(5- 22\fr{m_d}{m_s} \right) \left(\fr{1}{m^2_{H_1}}-\fr{1}{m^2_{\mathcal{A}}}\right)\right]. \eea Concerning the $B^0_{d,s}$-$\bar{B}^0_{d,s}$ mixing systems, we achieve similar expressions for $\Delta m_{B_d}$ and $\Delta m_{B_s}$ by replacing $(q'_i,q'_j)=(d,b)$ and $(s,b)$, respectively. Since $u\sim v$, the coefficient of $1/m^2_{H_1}-1/m^2_{\mathcal{A}}$ is significantly below that (i.e., 2/3) of $Z',Z''$. Even taking one of $u,v$ as small as $m_{K,B_{d,s}}\sim 1$ GeV, such coefficient is less than $\mathcal{O}(1)$, because of $v^2_{\mathrm{w}}m^2_{K,B_{d,s}}/12u^2v^2\sim 0.1$ and the associated factor $5-22m_{q'_i}/m_{q'_j}\sim 5$. The scalar contribution is strongly suppressed due to the $H_1,\mathcal{A}$ mass degeneracy, as ascertained above.            

\subsubsection{Remarks on natural flavor conservation principle}  

The 3-3-1-1 gauge symmetry by itself allows the soft term $f\epsilon^{ijk}\eta_i \rho_j \chi_k$ in the scalar potential. That said, the soft coupling $f$ naturally picks up a value to be the largest scale in the theory, $-f\sim (w,\La)$, since it is not suppressed by the symmetry. In this way, there is no tree-level FCNC coupled to the standard model Higgs boson. Additionally, although there exist tree-level FCNCs coupled to the new Higgs $H_1,\mathcal{A}$, their contributions to neutral meson mixing amplitude are cancelled out as strongly suppressed by $(u,v)^2/(w,\La)^2$, in similarity to the contributions of $H_{2,3}$ contained in $\chi_3,\phi$.

That said, there is no flavor-changing $t$-quark decay, such as $t\to c H$ and $t\to u H$, present in the model. Such processes also do not occur by emitting a new Higgs boson instead of $H$, since all the new Higgs fields have a mass at $w,\La$ scale beyond $t$ mass, given that the potential parameter $-f\sim (w,\La)$ as used throughout the text.

The 3-3-1-1 gauge principle as presented for suppressing dangerous FCNCs associated with scalars is indeed a realization/extension of a natural flavor conservation principle hypothesized long ago in \cite{gwfcnc}. The completion of the proof for FCNC suppression in the gauge sector can be found in our recent work \cite{ldtfcnc}.     

Last, but not least, requiring a tree-level FCNC coupled to the Higgs boson, as well as relevant flavor-changing top-quark decay phenomenology, necessarily violates the 3-3-1-1 suppression principle. The first work of Refs. \cite{fcnchiggs} would finetune the soft parameter $f$ to be low, somewhat as $f\sim (u,v)^2/w\sim 1$--10 GeV, at scale of the triplet scalar VEV in the type II seesaw mechanism. The second work of Refs. \cite{fcnchiggs} introduced a Peccei-Quinn symmetry to suppress the coupling $f \eta\rho\chi$ but allow it generated by a very large scalar, i.e. $f=\la_{\Phi} \langle \Phi\rangle $, where $\Phi$ carries a Peccei-Quinn charge broken by $\langle \Phi\rangle \sim 10^{10}$ GeV. But it is hard to understand an uncharacteristically-small value $\la_{\Phi}\sim 10^{-9}$--$10^{-10}$ (which obeys $f=1$--10~GeV), imposed in the mentioned work.        

\subsection{Numerical estimation}

\begin{figure}[h]
\bc
\includegraphics[scale=0.8]{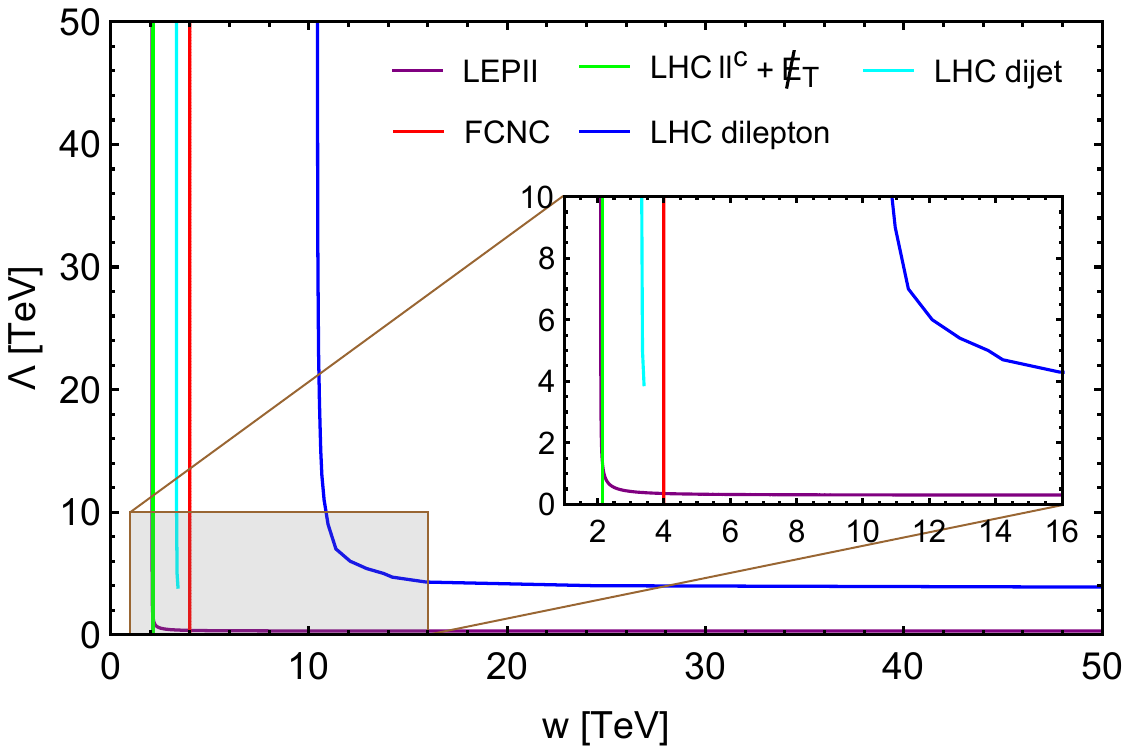}
\caption[]{\label{fig3} New physics scales $(w,\La)$ bounded by LEPII, LHC $ll^c+E\!\!\!\!/_T$, LHC dijet, FCNC and, LHC dilepton (corresponding curves arranged from left to right).}  
\ec
\end{figure}

As before, we take $s^2_W=0.231$, $\al=1/128$, and $t_G=1$, hence $t_X=\sqrt{3}s_W/\sqrt{3-4s^2_W}\simeq 0.577$ and $g_G=g=0.652$. It is clear from (\ref{mixingtheta}) and (\ref{zpzppmass}) that the $\mathcal{Z}'$-$C$ mixing angle $\theta$ and the $Z',Z''$ masses $m_{Z',Z''}$ depend only on the two new physics scales, $w,\La$. Hence, the constraints (\ref{lepc}) and (\ref{fcncc}) each directly yield a bound on $(w,\La)$, as despited in Fig. \ref{fig3}. Such a bound depends infinitesimally on $t_G$, i.e. the strength of the dark coupling $g_G$, if it varies. This is due to the fact that ordinary leptons and quarks have zero dark charge and the effects come only from small mixings. As already evaluated, when $\La$ is large, the FCNC is governed by $w$; conversely, when $\La$ is small, it does not contribute to FCNC, since $g'\to 0$. It follows that the FCNC bound is set by $w$ almost as the vertical line in the parameter space regime of interest, opposite to the LEPII that has a lower bound on $\La$. We also include in this figure (Fig. \ref{fig3}) the relevant bounds coming from LHC $ll^c+E\!\!\!\!/_T$, LHC dilepton, LHC dijet searches, as previously obtained. It is stressed that the LHC dilepton makes the strongest constraint on $(w,\La)$, even larger than the FCNC bound for $w$ and than the LEPII bound for $\La$, which is necessarily taken into account for neutrino mass and dark matter.   

To proceed further, the FCNC and collider constraints under consideration yield three distinct new physics regimes, such as  
\ben 
\item {\it 3-3-1 regime---the topmost regime in Fig. \ref{fig3}}: In the limit $\La\to \infty$ (or $\La\gg w$), we obtain a bound $w=10.422$~TeV by the LHC dilepton (radically bigger than the relevant FCNC bound $w=4$ TeV, as mentioned). In this case, $Z''$ is superheavy and decoupled from the 3-3-1 particle spectrum, while the $Z'$ mass is correspondingly limited by $m_{Z'}=4.135$ TeV. The 3-3-1 non-Hermitian gauge bosons $X,Y$ take a corresponding mass bound $m_{X,Y}\simeq (g/2)w\simeq 3.397$ TeV comparable to $Z'$, but larger than the LHC $ll^c+E\!\!\!\!/_T$ bound. All these $Z',X,Y$ bounds that are implied by the LHC for relevant 3-3-1 model have been well-established in the literature (see, e.g., \cite{futurelhc}).  
\item {\it Dark physics regime---the rightmost regime in Fig. \ref{fig3}}: In the limit $w\to \infty$ (or $w\gg \La$), we achieve a bound $\La=3.854$ TeV by the LHC dilepton (significantly larger than the relevant LEPII bound $\La=0.3$ TeV). In this case, $Z''$ and most of new particles are superheavy and decoupled from the standard model particle spectrum, except for the residual $U(1)_D$ symmetry and its relevant physics, whose $Z'$ dark gauge boson mass is correspondingly limited by $m_{Z'} = 3.388$ TeV. All these ingredients will be detailedly examined in the forthcoming section, Sec. \ref{darkcharge}.
\item {\it 3-3-1-1 regime---the rectangle regime in Fig. \ref{fig3}---as zoomed out for clarity}: In the case of $w\sim \La$, both $Z'$ and $Z''$ effectively govern the new physics. We fix benchmark values to be $(w,\La)=(12.088,6)$ or $(15.92,4.3)$, which translate to $(m_{Z'},m_{Z''})=(3.803,9.866)$ or $(3.415,10.37)$ respectively, where all values are given in TeV, following Table \ref{tab3}. This case belongs to the main interest of the work, which is subsequently studied in the rest of this section.  
\een

\begin{figure}[h]
\bc
\includegraphics[scale=1]{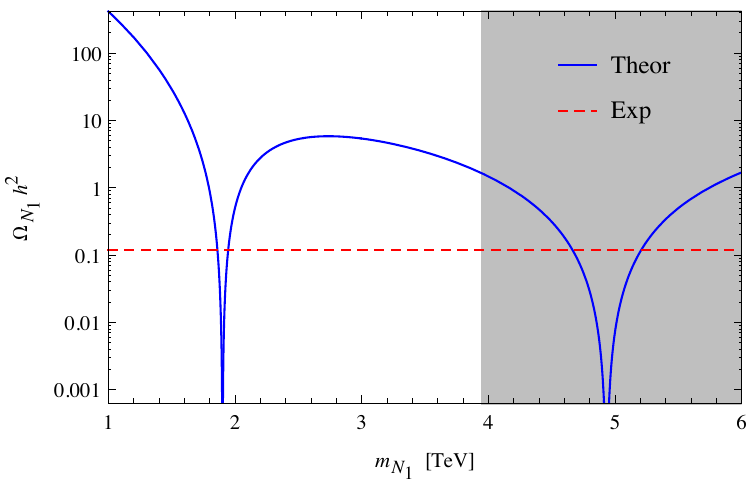}
\includegraphics[scale=1]{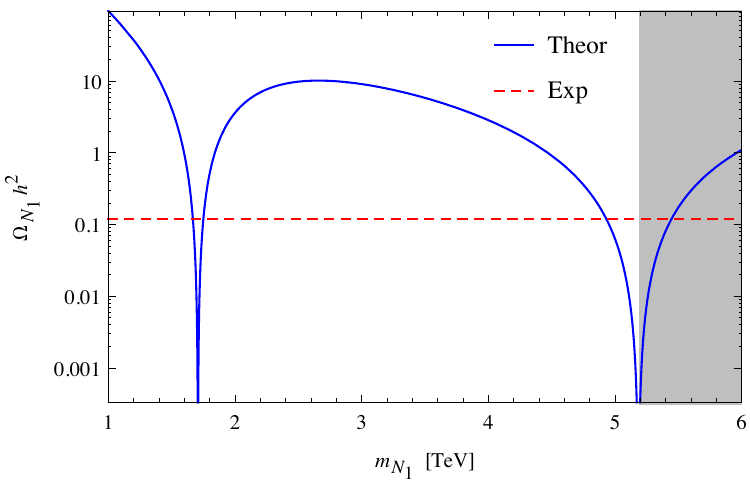}
\caption[]{\label{fig4} Dark matter relic density plotted as function of its mass according to two cases: $w=12.088$ TeV and $\La=6$ TeV (upper panel); $w=15.92$ TeV and $\La=4.3$ TeV (lower panel).}
\ec
\end{figure}

Using the parameter values and the last case, as given above, we plot the dark matter relic density (cf. Sec. \ref{dark}) as a function of the dark matter mass as in Fig. \ref{fig4} (solid curves).
 It is stressed that the $Z',Z''$ mass resonances (left, right funnels in each panel, respectively) are necessary to set the correct relic density, $\Om_{N_1}h^2\leq 0.12$ (dashed lines). For the case $(w,\La)=(12.088,6)$ TeV, the $Z'$ resonance $m_{N_1} = m_{Z'}/2$ plays the role, yielding $m_{N_1}=1.86$--1.95 TeV for the correct abundance, whereas the $Z''$ resonance is excluded by the WIMP unstable regime (shaded), namely $m_{N_1}<3.94$ TeV. However, for the case $(w,\La)=(15.92,4.3)$ TeV, both the resonances $m_{N_1}=m_{Z'}/2$ by $Z'$ and $m_{N_1}=m_{Z''}/2$ by $Z''$ take place. They indicate to $m_{N_1}=1.66$--1.75 TeV and $m_{N_1}=4.93$--5.19 TeV, for the correct abundance. Here note that the relic density is only satisfied for a part of the second resonance by $Z''$, since $m_{N_1}<5.19$ TeV ensuring WIMP stability, as limited below the shaded regime.  

\begin{figure}[h]
\bc
\includegraphics[scale=1]{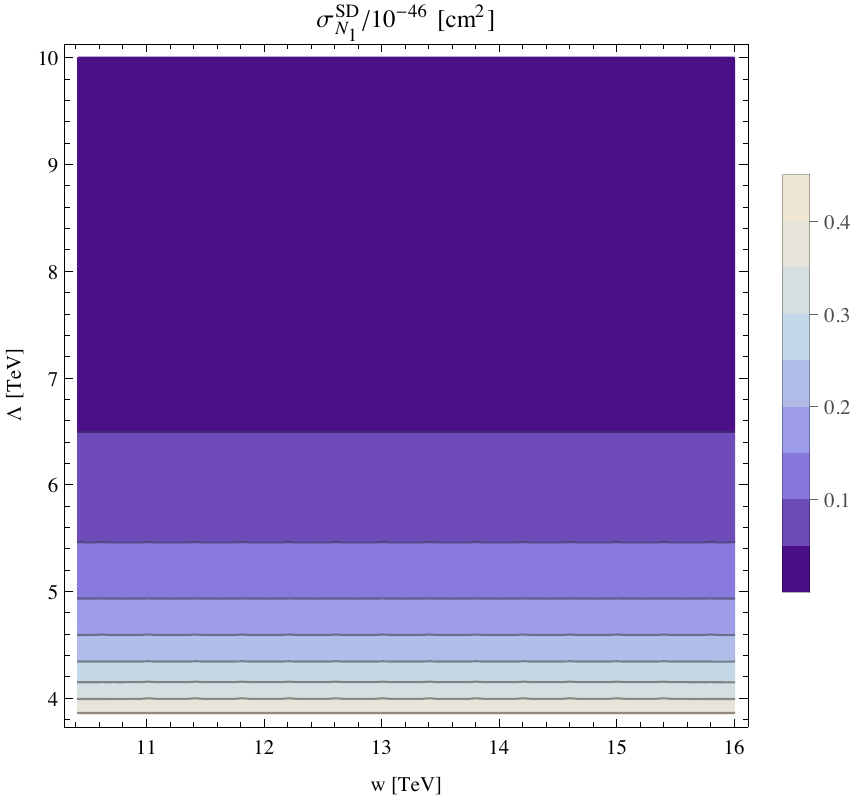}
\caption[]{\label{fig5} SD cross-section of $N_1$ contoured as a function of new physics scales $(w,\La)$ for $w\geq 10.422$~TeV and $\La\geq 3.854$~TeV.}
\ec
\end{figure}

With the aid of the limits obtained above for the new physics scales, i.e. $w> 10.422$ TeV and $\La>3.854$ TeV (cf. Fig. \ref{fig3}), as well as using the parameter values previously input for $s_W,\al,g_X,g_G$, we make a contour of the SD cross-section of dark matter with nuclei in direct detection experiment (cf. Sec. \ref{dark}) as a function of $(w,\La)$ as given in Fig. \ref{fig5}. 
 It is clear that the SD cross-section is more sensitive to $\La$ than $w$. Additionally, for viable regime $w\geq 10.422$ TeV and $\La\geq 3.854$ TeV, this model predicts the dark matter signal strength in direct detection to be $\sigma^{\mathrm{SD}}_{N_1}<10^{-46}\ \mathrm{cm}^2$, much below the current bound of $10^{-42}\ \mathrm{cm}^2$ order for a typical WIMP with mass beyond 1 GeV \cite{sdexp}.

\section{\label{darkcharge} Realization of the dark charge}

In this section, we consider an alternative scenario that reveals the main role of the dark charge by assuming the scalar triplet $\chi$ to be superheavy, possessing a VEV $w\gg \La$, and of course $\La\gg u,v$.\footnote{This case presents two new phases of the new physics similar to a matter discussed in \cite{sodl}.} Hence, the scheme of symmetry breaking is now 
\bc
\begin{tabular}{c}
$SU(3)_C \otimes SU(3)_L \otimes U(1)_X \otimes U(1)_G$\\
$\downarrow w $\\
$SU(3)_C \otimes SU(2)_L \otimes U(1)_Y \otimes U(1)_D$\\
$\downarrow \La $\\
$SU(3)_C \otimes SU(2)_L \otimes U(1)_Y \otimes P_D$\\
$\downarrow u,v $\\
$SU(3)_C \otimes U(1)_Q \otimes P_D$\\
\end{tabular} \ec
Indeed, when $\chi$ develops a VEV, $\langle \chi\rangle =(0,0,w/\sqrt{2})$, it breaks all new charges $T_{4,5,6,7,8}$, $X$, and $G$ but conserves $T_{1,2,3}$, $Y=-1/\sqrt{3}T_8+X$, and $D=-2/\sqrt{3}T_8+G$, besides the color, which match the standard model symmetry and $U(1)_D$, as expected. This breaking by $\chi$ decomposes every $SU(3)_L$ multiplet into a normal isomultiplet with $D=0$ and a dark isomultiplet with $D\neq 0$---known as dark isopartner of the normal isomultiplet---which all are possibly seen in Tab. \ref{tab1}. Given that the scale $w$ is very high, i.e. $w\gg \La\sim$ TeV, the new physics related to it, such as dark vectors $X,Y$ coupled to broken $T_{4,5,6,7}$, $Z''$ coupled to broken combination of $T_8,X,G$, relevant Goldstone bosons $G_X$, $G_Y$, and $G_{Z''}$ eaten by $X$, $Y$, and $Z''$ respectively, and its Higgs fields, is all decoupled/integrated out. What imprinted at scale $\La\sim$ TeV is a novel theory $SU(3)_C \otimes SU(2)_L \otimes U(1)_Y \otimes U(1)_D$, explicitly recognizing the dark charge $D$, directly affecting the standard model.

Notice that for $w\gg\La$, the $Z',Z''$ masses are \be  m^2_{Z'}\simeq \fr{4g^2_G(3+t_X^2)}{4t^2_G+3+t^2_X}\La^2,\hs m^2_{Z''}\simeq \fr{g^2}{9}(4 t^2_G+3+t^2_X)w^2,\ee and the $\mathcal{Z}'$-$C$ mixing angle is \be t_{\theta}\simeq \fr{\sqrt{3+t^2_X}}{2t_G}.\ee As mentioned, $Z''$ is decoupled, while $Z'$ associated with the dark charge now governs the collider signals, bounded by $m_{Z'}>3.388$~TeV for our choice of $t_G=1$ (see below in detail); additionally, the FCNC is suppressed as a result. In this case, $t_{\theta}\simeq 0.91$, i.e. $\theta\simeq 42.4^\mathrm{o}$, which determines the $Z'$ coupling with fermions, such as
\be \mathcal{L}\supset g_G s_\theta \sum_f \bar{f}\ga^\mu \left(-\fr 2 3 t^2_W Y + D\right) f Z'_\mu, \label{effzp}\ee where $f$ runs over usual lepton and quark isomultiplets as well as their dark isopartners. The presence of the $Y$ term like that from a kinetic mixing effect results from 3-3-1-1 breaking. That said, if the standard model fields have no dark charge $D=0$, they may interact with the dark boson $Z'$ through scotoelectroweak unification governed by the hypercharge $Y$. This effect is smaller than the dark force by one order, say $\fr 2 3 t^2_W\sim 0.1$.  

Although $\chi$ is superheavy, it can induce appropriate neutrino masses by the same mechanism and the result discussed above. But, the contribution of new physics in (\ref{numdd}) must be reordered, $(u/w)^2=(u/\La)^2\times (\La/w)^2\sim 10^{-3}\times 10^{-3}=10^{-6}$, the loop factor $(1/32\pi^2)\sim 10^{-3}$ as retained, the $N$ mass matrix being pseudo-Dirac such that $(h^N V)^2 M\sim (h^N \La/w)^2\times w=10^{-3}(h^N)^2w$, the scalar mass splitting as $\Delta m^2/m^2\sim (f_1,f_2 \la_{17})\La/w^2$. Hence, the neutrino masses are of order of eV,  
\be m_\nu\sim (h^N)^2\times \left(\fr{f_1,f_2\la_{17}}{w}\right)\times \left(\fr{\La}{\mathrm{TeV}}\right)\times  \mathrm{eV},\ee given that $h^N\sim 1$, $\La\sim $ TeV, and $f_{1,2}\sim w$, where the soft term ($f_{1,2}$) would mount to the scale of the 3-3-1-1 breaking.    

After decoupling by the large scale $w$, the intermediate TeV phase with $U(1)_D$ symmetry can contain some dark fields survived, such as $N_1$, $\xi$, and $\phi$ by choosing appropriate Yukawa couplings and scalar potential parameters. The dark matter phenomenology is similar to the above model, but it is now governed by only $Z'$ boson, coupled to normal matter via~(\ref{effzp}). For the dark fermion, the $Z'$ mass resonance sets its relic density. Alternatively, for the dark scalar, the new Higgs $\phi$ portal takes place annihilating to the standard model Higgs fields, since the dark scalar mass splitting in this case is large.           

\begin{figure}[h]
\bc
\includegraphics[scale=0.7]{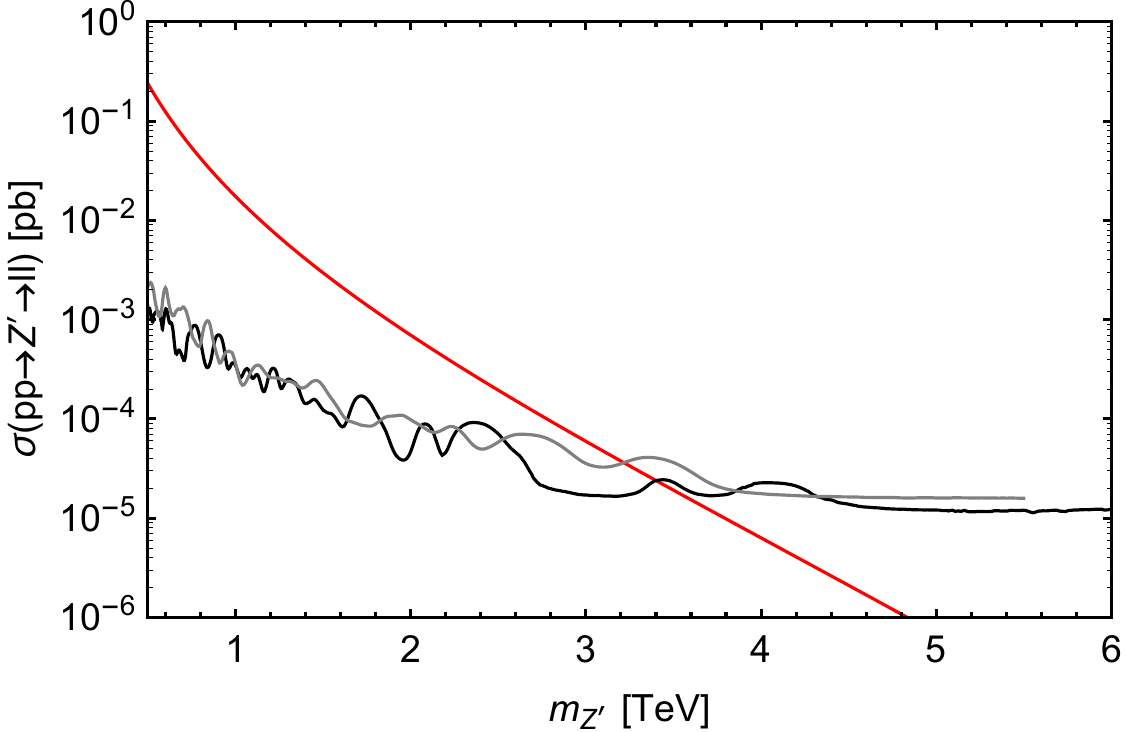}
\caption[]{\label{fig6} Dilepton production cross-section plotted as a function of $Z'$ dark boson mass at $pp$ collider for $\sqrt{s}=13$~TeV (red curve), where observed limits are extracted at dilepton invariant-mass resonance corresponding to ATLAS-2019 result for width $\Ga/m=3\%$ (black curve) \cite{atlascms1} and CMS-2021 result for width $\Ga/m=0.6\%$ (gray curve) \cite{atlascms2}.}
\ec
\end{figure}

Complementary to the LHC constraint, it is appropriate to verify the $Z'$ bound when the field $Z''$ is decoupled, i.e. $w\gg \La$, as above mentioned. Although this decoupling is taken, the result may apply for the case $w$ to be sufficiently separated from $\La$, i.e. relaxing $w$ raises beyond 50 TeV according to the third case in previous section for 3-3-1-1 model, such that $Z''$ negligibly contributes as compared to $Z'$ in the relevant LHC process. The cross section $\sigma(pp \to Z'\to l\bar{l})$ producing a dilepton final state $l\bar{l}$ at the LHC via $Z'$ exchange is already given by (\ref{lhcff}) for $f=l$ in narrow width approximation, in which the partonic peak cross-section $\hat{\sigma}(q\bar{q}\to Z')$ and the branching decay ratio $\mathrm{Br}(Z'\to l\bar{l})$ are given in (\ref{lhczp}) and (\ref{lhcdc}), respectively, too. Notice that the decay $Z'\to N_1 N_1$ insignificantly reduces the signal strength. We plot the dilepton production cross-section for $l=e,\mu,\tau$---which have the same couplings, thus production rate---as in Fig. \ref{fig6} at the LHC $\sqrt{s}=13$ TeV corresponding to an integrated luminosity of 139 $\mathrm{fb}^{-1}$ (ATLAS) \cite{atlascms1} and up to 140 $\mathrm{fb}^{-1}$ (CMS) \cite{atlascms2}. Both the ATLAS and CMS searches reveal a negative result for new dilepton event, hence making a bound for $Z'$ dark boson mass, such as $m_{Z'}>3.388$ TeV, which is significantly bigger than the LEPII limit at a few hundred GeV, as aforementioned. This translates to a limit for dark charge breaking scale, $\La=3.854$ TeV, as expected.

\section{\label{con} Conclusion}

The idea of a dark photon associated with a conserved, dark (abelian) charge is interesting as it provides potential solutions to a number of the current issues \cite{darkcharge0}. As electric charge is a result of electroweak breaking, this work has probed that a dark charge may result from a more fundamental theory, called scotoelectroweak. Moreover, the content of dark fields and the way they interact with normal matter are completely determined by the 3-3-1-1 symmetry of the theory.

We have examined the pattern of the 3-3-1-1 symmetry breaking, obtaining a residual dark parity that both stabilizes dark matter candidates and governs scotogenic neutrino mass generation. The small neutrino masses are suppressed by loop induced and ratio between electroweak to new physics scales, not requiring the soft terms to be too small. The fermion dark matter abundance is generically set by $Z',Z''$ mass resonances. Even in a scenario that the 3-3-1-1 breaking scale is very high, the light boson $Z'$ associated with the dark charge still plays the role due to a coupling to normal matter via the hypercharge. 

We have investigated the model under constraints from LEPII, LHC, and FCNCs. However, given a stronger bound it is easily evaded by enhancing $w,\La$ as the parameter space supplied in the figures. In all case, the signal for fermion dark matter in direct detection is very small. Embedding 3-3-1-1 symmetry in a GUT may be worth exploring as dark charge and its field contents may contribute to gauge coupling unification successfully.

\end{document}